\documentclass{article}
\usepackage{arxiv}
\usepackage{framed,multirow}
\usepackage{url}
\usepackage{xcolor}
\definecolor{newcolor}{rgb}{.8,.349,.1}
\usepackage{textcomp}
\usepackage[export]{adjustbox}
\usepackage{amsmath,graphicx}
\usepackage[utf8]{inputenc}
\usepackage{tabularx}
\usepackage{multirow}
\usepackage{interval}
\usepackage[lofdepth,lotdepth]{subfig}
\newcolumntype{L}[1]{>{\raggedright\let\newline\\\arraybackslash\hspace{0pt}}m{#1}}
\newcolumntype{C}[1]{>{\centering\let\newline\\\arraybackslash\hspace{0pt}}m{#1}}
\usepackage[colorlinks=true,citecolor=blue]{hyperref}
\usepackage{siunitx}
\usepackage{placeins}
\usepackage{graphicx}
\usepackage{multicol}
\usepackage{float}
\restylefloat{table}
\usepackage{hhline}
\usepackage{cite}
\usepackage{dsfont}
\usepackage{amssymb}
\newlength{\tempdima}
\newcommand{\rowname}[1]
{\rotatebox{90}{\makebox[\tempdima][c]{#1}}}

\title{Towards Histopathological Stain Invariance by Unsupervised Domain Augmentation using Generative Adversarial Networks}

\author{
 Jelica Vasiljevi\'{c} \\
  ICube, University of Strasbourg, CNRS (UMR 7357), France \\
  University of Belgrade, Belgrade, Serbia\\
  University of Kragujevac, Kragujevac, Serbia\\
  \texttt{jvasiljevic@unistra.fr} \\
  \And
 Friedrich Feuerhake \\
  Institute of Pathology, Hannover Medical School, Germany\\
  University Clinic, Freiburg, Germany\\
  \texttt{Feuerhake.Friedrich@mh-hannover.de}
  \And
 C\'{e}dric Wemmert \\
  ICube, University of Strasbourg, CNRS (UMR 7357), France\\
  \texttt{wemmert@unistra.fr}
  \And        
 Thomas Lampert \\
  ICube, University of Strasbourg, CNRS (UMR 7357), France\\
  \texttt{lampert@unistra.fr}
}

\begin{document}
\maketitle

\begin{abstract}
The application of supervised deep learning methods in digital pathology is limited due to their sensitivity to domain shift. 
Digital Pathology is an area prone to high variability due to many sources, including the common practice of evaluating several consecutive tissue sections stained with different staining protocols. Obtaining labels for each stain is very expensive and time consuming as it requires a high level of domain knowledge. In this article, we propose an unsupervised augmentation approach based on adversarial image-to-image translation, which facilitates the training of stain invariant supervised convolutional neural networks. By training the network on one commonly used staining modality and applying it to images that include corresponding, but differently stained, tissue structures, the presented method demonstrates significant improvements over other approaches. These benefits are illustrated in the problem of glomeruli segmentation in seven \color{black} different staining modalities (PAS, Jones H\&E, CD68, Sirius Red, CD34, H\&E and CD3) and analysis of the learned representations demonstrate their stain invariance.
\end{abstract}

\section{Introduction}

The introduction of Whole Slide Imaging (WSI) scanners enables the production of huge amounts of histological image data. A crucial step in the tissue preparation process is staining, which consists of dyeing the tissue slices with a specific stain in order to highlight internal structures when viewed under a microscope. Image datasets in digital pathology often consist of (nearly consecutive) tissue slides stained differently, each staining providing specific morphological information. As each staining highlights different components of the tissue, even consecutive sections largely representing almost identical anatomical structures can take on very different appearances. On the other hand, the staining process itself is prone to high variability due to inter-subject differences, the staining procedure in a specific lab, or scanner characteristics, leading to different appearances of the tissues coloured with the same stain  \cite{Leo2016EvaluatingSO}. This high variability in the appearance of digital histological images affects the performance of machine learning algorithms used for automatic WSI analysis \cite{Tellez2019QuantifyingTE}. In specific applications such as the detection or segmentation of objects that maintain morphological consistency across stains (e.g.\ nuclei, glomeruli, etc), stain variability could be considered, from a computer vision viewpoint, as noise that dramatically hampers the performance of machine learning models. This motivates the development of image analysis methods that work across different stains in order to solve tasks related to morphologically consistent structures \cite{Brieu2019DomainAA,gadermayrGenerativeAdversarialNetworks2019,gupta2018stain,lampert2019strategies}.

Most state-of-the-art computer vision algorithms are sensitive to domain shift \cite{csurka2017comprehensive}. This means that models trained for a specific task on histological images of stain A exhibit a significant drop in performance when applied to histological images of stain B (for the same task) \cite{Brieu2019DomainAA,gupta2018stain, lampert2019strategies} or variability of stain A (e.g.\ images from other laboratories) \cite{Tellez2019QuantifyingTE}.

Typical solutions consider either fine tuning existing models or training a new model for each data variation, which requires the acquisition of additional labeled data. Nevertheless, medical image datasets are often characterised by their scarcity of annotations \cite{EmbracingImperfectDatasets} and obtaining properly annotated images is time consuming and costly 
\color{black}as expert knowledge is required for most complex annotations. Alternative approaches such as crowd sourcing are limited by the need of specific task design and intensive training \cite{grote2018crowdsourcing}. 
\color{black} While plenty of works address the problem of intra-stain variation \cite{Bel2019StainTransformingCG,Bentaieb2018AdversarialST,Lafarge2017DomainAdversarialNN,Lafarge2019LearningDR,Macenko09,Otlora2019StainingIF,Shaban2019StainganSS} there are few that address the problem of inter-stain variation \cite{Brieu2019DomainAA, gadermayrGenerativeAdversarialNetworks2019,Lahiani2018VirtualizationOT,lampert2019strategies,Mercan2020VirtualSF}.

Common approaches to deal with inter-stain variability are stain colour augmentation, stain normalisation, and stain transfer \cite{Srinidhi2019DeepNN}. A recent study on the effects of colour augmentation and stain normalisation for computational pathology shows that stain colour augmentation has greater impact on the robustness of machine learning models compared to stain normalisation \cite{Tellez2019QuantifyingTE}. On the other hand, stain transfer as a technique for virtual staining \cite{Lahiani2018VirtualizationOT,Mercan2020VirtualSF} addresses the general problem of a lack of annotations in the medical domain and can be applied to various problem classes.

In this article we improve upon existing work \cite{lampert2019strategies} to introduce a general approach that combines stain transfer and stain augmentation to learn stain invariant representations\footnote{\color{black}‘Stain invariant’ refers to the fact that the same model can be applied to multiple stains (possibly those not seen during training) without modification of the data at test time or adaptation of the model after training.}.
Generally, it is assumed that annotations exist for one staining (called the source staining) and that other stainings (called the target stainings) are unannotated. The goal is to use the available annotations to learn a stain invariant representation by taking advantage of stain transfer techniques.

We focus on the segmentation of glomeruli, a highly relevant functional unit of the kidney, and consider a commonly used staining in renal pathology---Periodic acid-Schiff reaction (PAS)---as the source domain and two commonly applied histochemical (Sirius Red and Jones' basement membrane stain---Jones H\&E) and two immunohistochemical stainings (CD34 highlighting blood vessel endothelium, CD68 for macrophages) as targets.

Specifically, we train GAN-based stain-translation models \cite{StarGAN2018,CycleGAN2017} in order to achieve realistic translation between source and target stains. 

In turn, these allow a segmentation model to be trained on source data augmented with random translations to the target stains. Compared to previous state-of-the-art approaches to stain-invariant segmentation \cite{gadermayrGenerativeAdversarialNetworks2019,lampert2019strategies}  the presented strategy demonstrates a significant improvement in segmentation performance.

The main contributions of this article are:
\begin{itemize}
\item to introduce a model that is capable of segmenting several stainings, i.e.\ is stain invariant, and outperforms all evaluated state-of-the-art approaches;
\item an augmentation method that is general and unsupervised;
\item to the best of our knowledge, this is the first approach to use image-to-image translation to force stain-invariant features to be learnt, rather than using the typical domain adaptation process that results in domain (staining) specific models, albeit different from the source staining;
\item to increase flexibility regarding the direction of translation described in the literature \cite{gadermayrGenerativeAdversarialNetworks2019};
\item to present a discussion on the limitations of GAN-based image-to-image translations. \color {black} 
\end{itemize}

The remainder of this article is organised as follows: in Section \ref{sec:Related_work}, literature concerned with stain transfer and stain invariant approaches is reviewed. Section \ref{sec:Method} gives a detailed description of the proposed method and dataset. Section \ref{sec:Results} presents the experimental results with comparison to existing state-of-the-art methods. Finally, Section \ref{sec:Discussion} analyses the approach in terms of stain invariance, quality of translations, and its limitations.

\begin{figure}
	\begin{center}
\small{
		\settoheight{\tempdima}{\includegraphics[width=1.6cm]{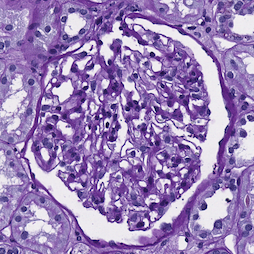}}%
		\begin{tabular}{@{}c@{ }c@{ }c@{ }c@{ }c@{ }c@{ }}
			& PAS & Jones H\&E  & CD68 & Sirius Red & CD34
			\\
			\rowname{\small{Original}} & 
			\includegraphics[width=1.6cm]{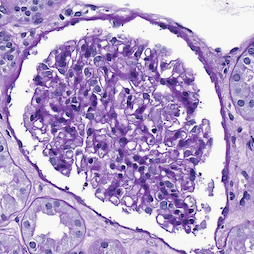} &
			\includegraphics[width=1.6cm]{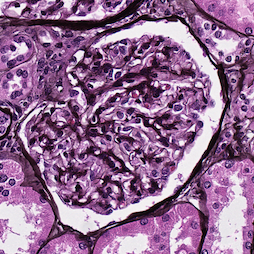} &
			\includegraphics[width=1.6cm]{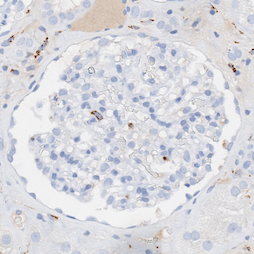} &
			\includegraphics[width=1.6cm]{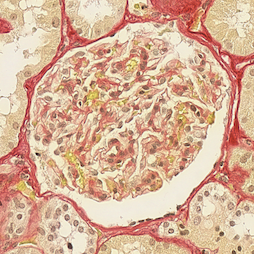} &
			\includegraphics[width=1.6cm]{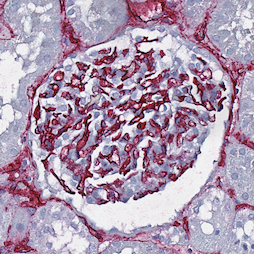}
			\vspace{0.8ex}
			\\
			\multirow{2}{*}{\rowname{Translation to Targets}}
			& \rowname{{\small CycleGAN}}
			  &
			\includegraphics[width=1.6cm]{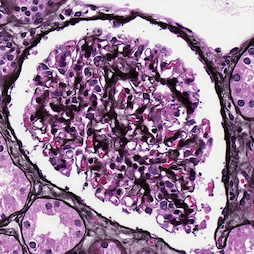} &
			\includegraphics[width=1.6cm]{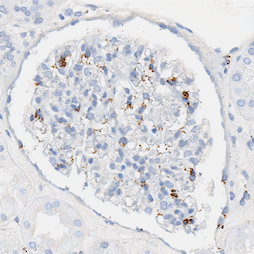} &
			\includegraphics[width=1.6cm]{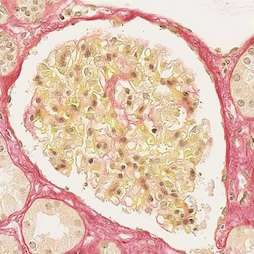} &
			\includegraphics[width=1.6cm]{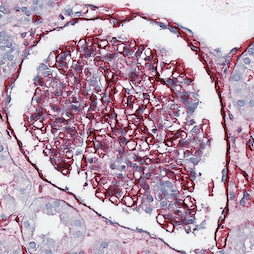}
			\\
			 & \rowname{{\small StarGAN}}&
			\includegraphics[width=1.6cm]{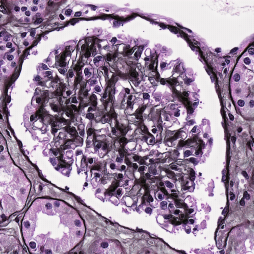} &
			\includegraphics[width=1.6cm]{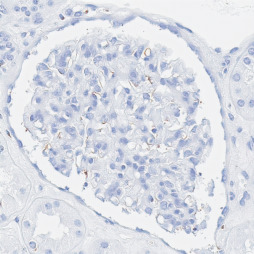} &
			\includegraphics[width=1.6cm]{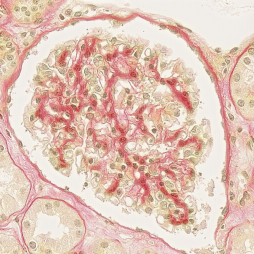} &
			\includegraphics[width=1.6cm]{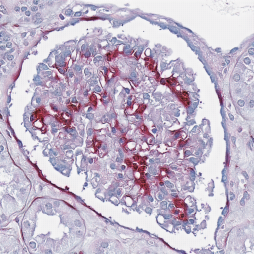}
			\vspace{0.8ex}
			\\
			\multirow{2}{*}{\rowname{Translation to PAS}} &
			\rowname{{\small CycleGAN}}& 
			\includegraphics[width=1.6cm]{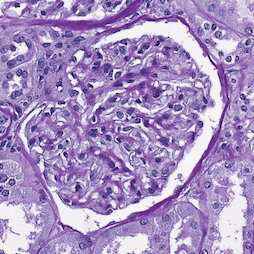} &
			\includegraphics[width=1.6cm]{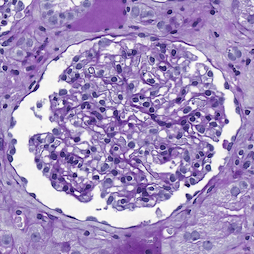} &
			\includegraphics[width=1.6cm]{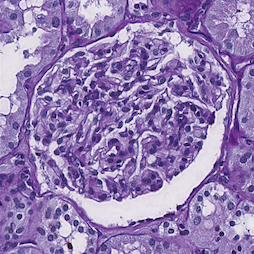} &
			\includegraphics[width=1.6cm]{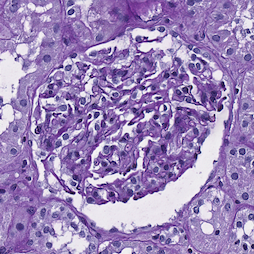}
			\\
			&  \rowname{{\small StarGAN}}& 
			\includegraphics[width=1.6cm]{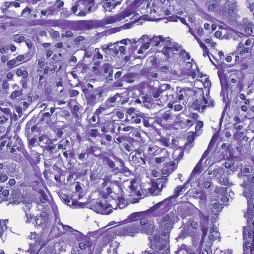} &
			\includegraphics[width=1.6cm]{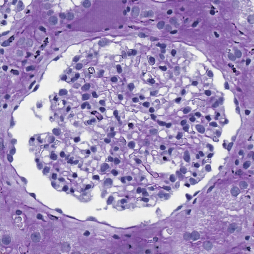} &
			\includegraphics[width=1.6cm]{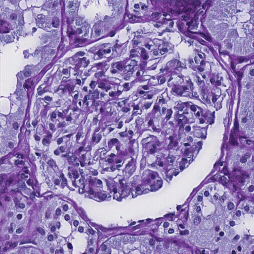} &
			\includegraphics[width=1.6cm]{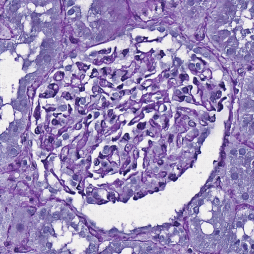}
		\end{tabular}
		}
		\caption {Examples of stain-to-stain translated images using a CycleGAN. The first row contains real patches from each staining. The second row represents the translation of a PAS patch to the target stainings. The third row represents translation of real target patches to the PAS staining.}
		\label{fig:cyclegan_translation}

	\end{center}
\end{figure}


\section{Related work}
\label{sec:Related_work}

The occurrence of stain variability can be grouped into intra-stain variability (variations in appearance of a particular stain) or inter-stain variability (the variability in appearance due to different stains). As has been shown in the literature, both types of variation hamper the performance of deep models \cite{lampert2019strategies,Tellez2019QuantifyingTE}. 
When the task at hand is related to the morphological structures in the tissue (such as glomeruli or nuclei detection) rather than stain-related markers (such as macrophages) approaches invariant to inter-stain variation are highly desirable.

Increasing a model's robustness to general stain variations can be considered a domain adaptation problem, which can be approached in either pixel-space or feature space. 
In feature space, the adaptation is performed by forcing the extraction of domain invariant features during model training. 
In pixel space, the adaptation can be performed by modifying the properties of a target image in order to match characteristics of the source images, so models trained on the annotated domain can be applied to the modified target images. Conversely the annotated domain can be translated to match the non-annotated (in a way that annotations are preserved), and train a model on the transformed source images such that it is directly applicable in the target domain. Adaptation in pixel space offers better interpretation of the model (from a end-user, i.e.\ clinician or pathologist, point of view), contrary to mapping them in a feature space, since the result can be viewed as a histopathological image. This is an important consideration in medical imaging. Moreover, once the pixel-space has been aligned, any task can be performed within it.

Inter-stain variability approached from the perspective of feature space alignment is addressed in recent works \cite{DualAdaptivePyramidNetwork,CrossStainedSegmentation}. 

In general, feature space alignment has proven effective in transferring knowledge from single (or multiple) source domains to a single target domain, i.e.\ producing a task and stain specific model. Recently, advances move towards single-source multi-target approaches \cite{Gholami20}, however they still rely on task specific features. Most of the work on unsupervised domain adaption (i.e.\ no labels in the target domain) are single-source single-target approaches, which fail to leverage knowledge shared across multiple domains.

From a pixel-space alignment perspective, stain colour augmentation, stain normalisation and stain transfer are commonly used to prevent models from being biased towards the source stain \cite{Srinidhi2019DeepNN}. Stain normalisation refers to the standardisation of a stain's visual appearance and is commonly used as a pre-processing technique to reduce intra-stain variability \cite{Bel2019StainTransformingCG,Bentaieb2018AdversarialST,Shaban2019StainganSS}. Stain colour augmentation and stain transfer are broader techniques that can address inter-stain variability.

Stain augmentation approaches aim to synthesise new images to enforce colour variation robustness, assuming that objects of interests are invariant to colour intensity and illumination. A typical approach is to perturb image pixels in a random or weighted manner \cite{lampert2019strategies,Tellez18,XiaoColorAugmentation2019}. Alternatively, the stain variation problem can be bypassed by using grayscale images but several studies show this to be inferior \cite{lampert2019strategies,Tellez2019QuantifyingTE}.

Stain transfer as a virtual staining technique aims to translate an image of tissue stained with a particular stain (or even unstained tissue \cite{Bayramoglu2017TowardsVH,Rana2018ComputationalHS}) into another stain as realistically as possible \cite{Lahiani2018VirtualizationOT,Mercan2020VirtualSF,Xu2019GANbasedVR}. Earlier methods
\cite{Macenko09,Reinhard01,Vahadane16} rely on decomposing the image into concentration and colour matrices in order to stain new images using equivalent matrices form a reference image. Despite the use of a reference image, these approaches suffer from poor translational realism \cite{lampert2019strategies}. Recently, GANs \cite{goodfellow2014generative} have been used for this purpose due to their ability to synthesize realistic outputs. These are applied in two settings: unpaired translations \cite{gadermayrGenerativeAdversarialNetworks2019,Xu2019GANbasedVR} and paired translation \cite{Mercan2020VirtualSF}. The CycleGAN architecture \cite{CycleGAN2017}, as a widely used technique for unpaired image-to-image translation, has recently been applied to stain transfer \cite{Brieu2019DomainAA,gadermayrGenerativeAdversarialNetworks2019,Mercan2020VirtualSF, Xu2019GANbasedVR}. And StarGAN \cite{StarGAN2018}, a technique of multi domain image-to-image translation, has recently been applied to virtual histological staining of bright-field microscopic images \cite{LiDan2020}.

Gadermayr et al.\ \cite{gadermayrGenerativeAdversarialNetworks2019} showed that, when applied to the task of histological stain transfer, CycleGAN is able to produce highly realistic translations, increasing a deep model's robustness in both supervised and unsupervised settings. They proposed three approaches to overcome a lack of target domain annotations \cite{gadermayrGenerativeAdversarialNetworks2019}: 1) train a segmentation model on source data and apply it to target data translated to the source domain, referred to as MultiDomain Supervised 1 (MDS1); 2) train on the target stain translated to the source, and apply the model directly on the target images, MDS2.
Both MDS1 and MDS2 approaches build stain-specific models that can be applied only to a specific staining. \color{black} The third approach directly trains the segmentation model by optimising unpaired image-to-segmentation translations in an adversarial manner \cite{gadermayrGenerativeAdversarialNetworks2019}. However, due to the specificity of cycle-consistency, this approach requires that additional information be artificially generated and inserted into the annotation (e.g.\ cell nuclei). Furthermore, the performance of such an approach was consistently surpassed by the MDS approaches.

The work proposed herein represents a fourth approach that uses the principles of pixel-space alignment to achieve stain transfer and to form a stain-invariant model capable of direct application to several stains. The presented method is general, unsupervised, and applicable to any problem that concerns training a segmentation/classification model for problems that tackle morphologically consistent structures in digital pathology. The only existing similar approach is UDA-SD \cite{lampert2019strategies}, which we build upon by using GANs to obtain high-quality translations between annotated and unannotated stains (see Figure \ref{fig:cyclegan_translation}). These are used to augment a training set by random translations of the source domain, resulting in one stain-invariant model capable of segmenting various (unlabeled) domains. As such, this work proposes an single-source, multi-target unsupervised domain adaptation approach.

\begin{figure*}
	\centering
	\includegraphics[width=0.8\textwidth]{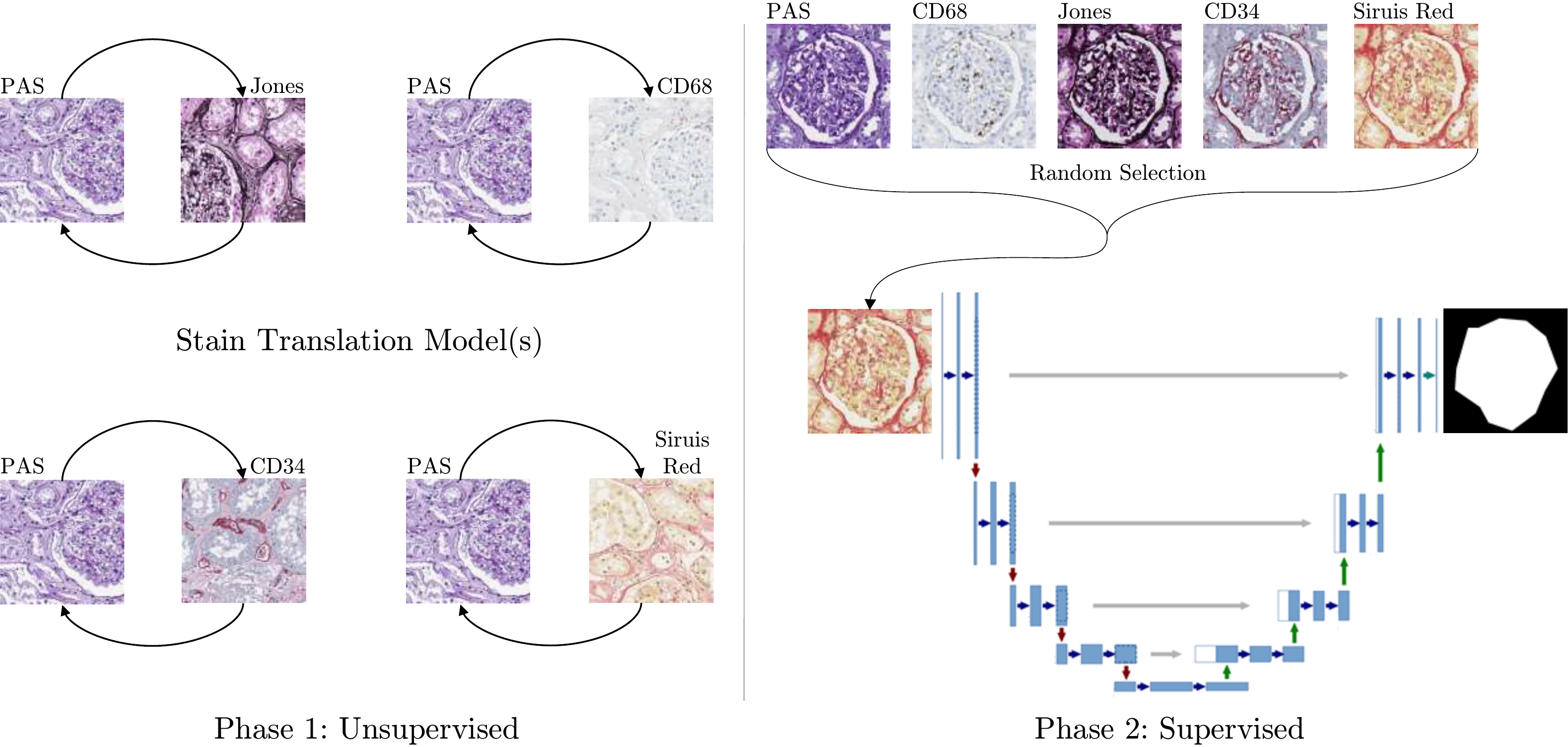}
	\caption{Overall diagram of the proposed approach. Phase 1, translation models are learnt to translate images from the source domain to the target domains; Phase 2, patches of the source domain are randomly translated to the target domains during training (U-Net image taken from \cite{unet}).}
	\label{fig:proposed_approach}
\end{figure*}


\section{Method}
\label{sec:Method}

We propose an approach for training stain invariant Convolutional Neural Networks (CNNs). It is assumed that annotated WSIs are available for a stain $A$ while WSIs of other stains $B_1, B_2,\dots,B_N$ are unannotated. The aim is to increase the variability of the (annotated) training set through augmentation by randomly translating it to the non-annotated domains (including the original, annotated domain). The overall architecture of the proposed method is presented in Figure \ref{fig:proposed_approach}.

The method consists of the following steps.

\setcounter{secnumdepth}{5}

\begin{figure*}[tb!]
	\begin{center}
	\subfloat[]{
		\centering
		\includegraphics[width=0.54\textwidth,valign=c]{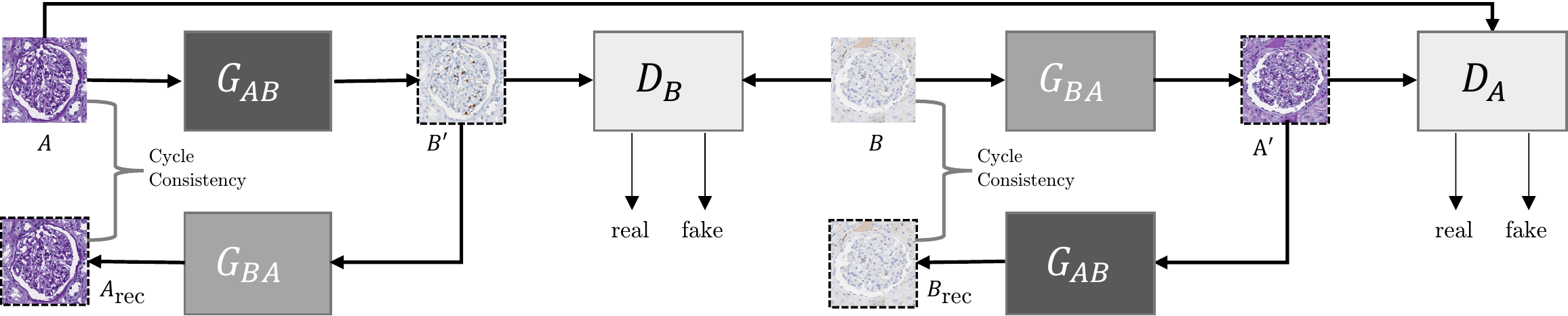}
		\label{fig:cyclegan}
	} 
	\subfloat[]{
		\centering
		\includegraphics[width=0.42\textwidth,valign=c]{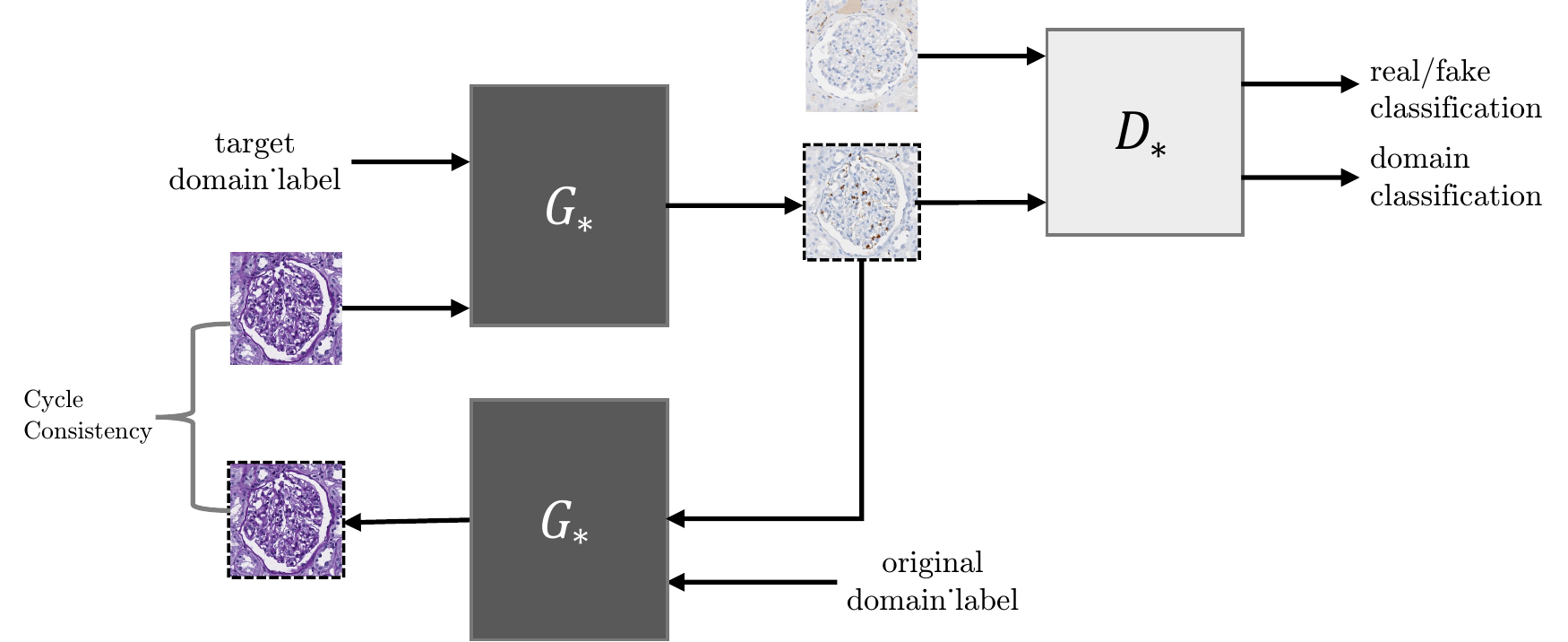}
		\label{fig:stargan}
	}
		\caption{\color{black}CycleGAN (a) and StarGAN (b) diagrams.\color{black}}
		\label{fig:stain_transfer_diagrams}
	\end{center}
\end{figure*}

\emph{a) Stain Translation} - \color{black} in order to obtain realistic translations of the annotated stain $A$ to unannotated stains $B_1, B_2,\dots,B_N$, a GAN-based unsupervised, unpaired image-to-image translation model is employed. In this study we evaluate both the CycleGAN \cite{CycleGAN2017} and StarGAN \cite{StarGAN2018} models (see Figure \ref{fig:stain_transfer_diagrams}).
Once these translation models are obtained, any supervised model (for which labels exist in the source domain and tackles morphologically consistent structures) can be trained. This study focuses on glomeruli segmentation.

\emph{b) Segmentation Model} - this model is trained on the annotated data after being translated to a random non-annotated stain. Since translation does not change the overall structure of the image (Figure \ref{fig:cyclegan_translation}), the original domain's label/ground truth is still valid. Thus, during training, various annotated samples of all available stainings are presented to the model, forcing it to learn stain invariant features. Once the segmentation model has been trained, it can be directly applied to the unannotated stains, without any further translations.

\subsection{Stain Translation Model}
\label{sec:StainTranslationModel}
The problem of stain transfer is posed as an unpaired image-to-image translation, which is approached using GAN-based models. The overall structure of these approaches assume the existence of one (or more) generators that translate an image from one domain to another, and one (or more) discriminators that distinguish between translated and real images from corresponding domains. The generator(s) and discriminator(s) play an adversarial game in which the goal of the generator(s) is to prevent the discriminator(s) from distinguishing between the translated samples and those belonging to the real data distribution, while the goal of the discriminator(s) is to distinguish them. Eventually, this game leads to an equilibrium in which the translated and real samples are indistinguishable \cite{StarGAN2018,goodfellow2014generative,CycleGAN2017}. Since the task of image-to-image translation is usually applied to natural images, the visual quality and variation of translated samples are very important. Thus, various architectures have been proposed in order to obtain better quality translations and greater style transfer capabilities (e.g.\ StarGAN v2 \cite{StarGAN2018_v2}). 

Nevertheless, since the purpose of stain transfer is to translate an image of stain $A$ to have the appearance of stain $B$, it is necessary for the translation model to preserve the image's global structure (e.g.\ glomeruli should remain in the same location, regardless of the target stain). Thus, image-to-image translation architectures which are less constrained and allow more diverse translation (including shape change) such as StarGAN v2 \cite{StarGAN2018_v2} are not suitable. 
The specific architecture used in the CycleGAN and StarGAN models, combined with the loss functions (particularly cycle-consistency) in practice prevents geometrical changes \cite{CycleGAN2017}, as is attested by the number of prior works using CycleGANs in digital pathology  \cite{Bel2019StainTransformingCG,Brieu2019DomainAA,gadermayr2018which,gadermayrGenerativeAdversarialNetworks2019,Lahiani2018VirtualizationOT,Mercan2020VirtualSF,Shaban2019StainganSS}.

One such example is given by \cite{gadermayr2018which}, who uses the CycleGAN for realistic stain-transfer. 
StarGAN, however, has thus-far not been used for this purpose and we observe the same geometrical consistency when applied to stain transfer.

\subsubsection{CycleGAN Based Stain Transfer}

Separate CycleGAN models, see Figure \ref{fig:cyclegan}, are trained to obtain translators between each target stain ($B$) and the source stain ($A$). Each model consists of two generators: $G_{AB}: A\rightarrow B$ and $G_{BA}: B\rightarrow A$; and two discriminators $D_A$ and $D_B$. The aim of $D_A$ is to distinguish between real source stain patches and those translated from the target stain to the source; while $D_B$ aims to distinguish between real target stain patches and those translated from the source stain to the target.
These are trained in an adversarial manner \cite{goodfellow2014generative}. 

\subsubsection{StarGAN Based Stain Transfer}
\color{black}
Rather than training a translation model for each pair of stains, StarGAN \cite{StarGAN2018} results in one, multi-stain translation model, see Figure \ref{fig:stargan}. This model contains one generator $G_*$, conditioned on the domain label (stain), which translates an input image $x_i$ from stain $i$ to image $x_j$, having the characteristics of stain $j$, i.e.\ $G_*(x_{i},j)\rightarrow x_{j}$; and one multi-task discriminator $D_*$ that simultaneously distinguishes between real and generated samples ($D_{adv}$) and classifies each patch's stain label ($D_{stain}$), i.e.\ $D_*(x)\rightarrow (D_{stain}(x),D_{adv}(x))$. As such, a single discriminator controls the translation into multiple stains.

\subsection{Segmentation Model}
\label{sec:SegModel}

The U-Net \cite{unet} is adopted as it has been proven successful in biomedical imaging \cite{litjens2017survey} and, in particular, glomeruli detection \cite{de2018automatic}. Glomeruli segmentation is framed as a two class problem: glomeruli (pixels that belong to glomerulus),  and tissue (pixels outside a glomerulus).

During training, a training patch with a probability of $\frac{N-1}{N}$, where $N$ is the number of stains, is converted into randomly selected stain using the pre-trained translator(s). Thus, all available stains (including the annotated one) are presented to the network with equal probability $\tfrac{1}{N}$, forcing the network to learn more stain invariant features

\subsection{Training Details}
Throughout the study, patches of size $508 \times 508$ pixels are used  since glomeruli and part of the surrounding fit within this size of patch at the level-of-detail used, see Section \ref{sec:Data}.

\subsubsection{Stain translation}
 Gadermayr et al.\ \cite{gadermayr2018which} showed that different sampling strategies for the annotated and unannotated domains could negatively impact a stain transfer model's performance, and patches are therefore randomly extracted using a uniform sampling strategy (in an unsupervised manner).
 
\noindent{\emph{CycleGAN: }}
In order to stabilise adversarial training, CycleGAN \cite{CycleGAN2017} uses least-squares loss instead of the original negative log-likelihood, such that

\begin{flalign}
\mathcal{L}_{\text{CycleGAN}}(G_{AB},G_{BA},D_A,D_B) =  \mathcal{L}_{\text{adv}}(G_{AB},D_B) \notag\\
									+ \mathcal{L}_{\text{adv}}(G_{BA},D_A)\notag
									+ w_{\textrm{cyc}} \mathcal{L}_{\text{cyc}}(G_{AB},G_{BA})\notag\\
									+ w_{\textrm{id}} \mathcal{L}_{\text{identity}}(G_{AB},G_{BA}),
\label{eq:CycleGAN_loss}
\end{flalign}
where
\begin{equation*}
\begin{split}
\mathcal{L}_{\text{adv}}(G_{AB},D_B) &= \mathds{E}_{t \sim P_{T_i}(t)}[(D_B (t) -1)^2]\notag\\&+ \mathds{E}_{s \sim P_{S}(s)} [(D_B(G_{AB}(s))-1)^2],
\end{split}
\end{equation*}
\begin{equation*}
\begin{split}
\mathcal{L}_{\text{adv}}(G_{BA},D_A) &= \mathds{E}_{s \sim P_{S}(s)} [(D_A(s)-1)^2]\notag\\&+ \mathds{E}_{t \sim P_{T_i}(t)} [(D_A(G_{BA}(t))-1)^2],
\end{split}
\end{equation*}
\begin{equation*}
\begin{split}
\mathcal{L}_{\text{cyc}}(G_{AB},G_{BA}) &= \mathds{E}_{s \sim P_{S}(s)} [\| G_{BA}(G_{AB}(s))-s \|_1]\\&+ \mathds{E}_{t \sim P_{T_i}(t)} [\| G_{AB}(G_{BA}(t))-t \|_1],
\end{split}
\label{eq:cycle}
\end{equation*}
\begin{equation*}
\begin{split}
\mathcal{L}_{\text{identity}}(G_{AB},G_{BA}) &= \mathds{E}_{s \sim P_{S}(s)} [\| G_{BA}(s)-s \|_1]\notag\\&+ \mathds{E}_{t \sim P_{T_i}(t)} [\| G_{AB}(t)-t \|_1].
\end{split}
\end{equation*}
The loss weights and architecture are taken from the original paper ($w_{\textrm{cyc}}=10$, $w_{\textrm{id}}=5$) \cite{CycleGAN2017} since they produced realistic output, see Fig.\ \ref{fig:cyclegan_translation}. A translation network with nine ResNet blocks was used, as suggested by the authors for high dimensional data (above $256 \times 256$ pixels) \cite{CycleGAN2017}.

The models are trained for $50$ epochs, with a learning rate of $0.0002$, using the Adam optimiser, and a batch size of $1$. From the $25^\textrm{th}$ epoch, the learning rate linearly decayed to $0$, and the cycle-consistency and identity weights halved. In all experiments, the translation model from last ($50^\textrm{th}$) epoch is used. 
Moreover, to reduce model oscillation, Shrivastava et al.’s strategy \cite{Shrivastava17} of updating the discriminator using the $50$ previously generated samples is adopted.

\noindent{\emph{StarGAN: }}

In order to promote adversarial training stability, StarGAN \cite{StarGAN2018} uses Wasserstein loss with gradient penalty instead of the original negative log-likelihood, such that 

\begin{flalign*}
\mathcal{L}_{\text{StarGAN}}(G_*,D_*) =  \mathcal{L}_{\text{adv}}(G_*,D_*) \notag
+ w_{\textrm{cyc}} \mathcal{L}_{cyc}(G_*)\notag\\
+ w_{\textrm{cls}} \mathcal{L}_{cls}(G_*, D_*),
\label{eq:StarGAN_loss}
\end{flalign*}
where
\color{black}
\begin{equation*}
\begin{split}
\mathcal{L}_{\text{adv}}(G_*,D_*) &= \mathds{E}_{x \sim P_j(x)}[D_{\text{adv}}(x)]\notag\\
&- \mathds{E}_{(x \sim P_i(x),j)} [D_{\text{adv}}(G_*(x,j)))] \\& - \lambda_{gp} \mathds{E}_{\hat{x}}[(\| \nabla_{\hat{x}}D_{adv}(\hat{x})\|_2 -1)^2],
\end{split}
\end{equation*}
\begin{equation*}
\mathcal{L}_{\text{cyc}}(G_*) = \mathds{E}_{x \sim P_i(x),i,j} [\| G_*(G_*(x,j),i)-x \|_1],
\end{equation*}
\begin{equation*}
\begin{split}
\mathcal{L}_{\text{cls}}(G_*,D_*) &=  \mathds{E}_{(x \sim P_i(x), j)}[- \log D_{\text{stain}}(j|G_*(x_i,j))]\notag\\ 
&+ \mathds{E}_{(x \sim P_i(x))}[- \log D_{\text{stain}}(i|x)],
\end{split}
\end{equation*}
where $\hat{x}$ is sampled uniformly between the real and generated images \cite{StarGAN2018}.

The generator's and discriminator's architecture and training settings were the same as the CycleGAN model described previously (without instance normalisation in the discriminator). The loss weights are taken from the original paper ($w_{\textrm{cyc}}=10$, $w_{\textrm{cls}}=1$, $\lambda_{gp}=10$) \cite{StarGAN2018} since they produced realistic output, as can be seen in Figure \ref{fig:cyclegan_translation}. 

\subsubsection{U-Net}
The same training parameters are used for all experiments: batch size of $8$, learning rate of $0.0001$, $250$ epochs, and the network with the lowest validation loss is kept.

The slide background (non-tissue) is removed by thresholding each image by its mean value then removing small objects and closing holes. 

All patches are standardised to $\interval{0}{1}$ and normalised by the mean and standard deviation of the (labeled) training set. 

After translation (or not), the following augmentations are applied with an independent probability of $0.5$ (batches are augmented `on the fly'), in order to further force the network to learn general features: elastic deformation  ($\sigma = 10$, $\alpha = 100$); random rotation in the range $\interval{\ang{0}}{\ang{180}}$, random shift sampled from $\interval{-205}{205}$ pixels, random magnification sampled from $\interval{0.8}{1.2}$, and horizontal/vertical flip; additive Gaussian noise with $\sigma \in \interval{0}{2.55}$; Gaussian filtering with $\sigma \in \interval{0}{1}$; brightness, colour, and contrast enhancements with factors sampled from $\interval{0.9}{1.1}$; stain variation by colour deconvolution \cite{Tellez18}, $\alpha$ sampled from $\interval{-0.25}{0.25}$ and $\beta$ from $\interval{-0.05}{0.05}$.
\subsection{Data}
\label{sec:Data}
Tissue samples were collected from a cohort of $10$ patients who underwent tumor nephrectomy due to renal carcinoma. The kidney tissue was selected as distant as possible from the tumors to display largely normal renal glomeruli, some samples included variable degrees of pathological changes such as full or partial replacement of the functional tissue by fibrotic changes (``scerosis'') reflecting normal age-related changes or the renal consequences of general cardiovascular comorbidity (e.g.\ cardial arrhythmia, hypertension, arteriosclerosis). The paraffin-embedded samples were cut into $3 \mu m$ thick sections and stained with either Jones' H\&E basement membrane stain (Jones), PAS, Sirius Red or H\&E, in addition to three immunohistochemistry markers (CD34, CD68 and CD3), using an automated staining instrument (Ventana Benchmark Ultra). H\&E and CD3 were reserved as unseen stain, therefore the total number of augmentation stains is $N = 5$. Whole slide images were acquired using an Aperio AT2 scanner at $40\times$ magnification (a resolution of 0.253 $\mu m$ / pixel). All the glomeruli in each WSI were annotated and validated by pathology experts by outlining them using Cytomine \cite{maree2016collaborative}. The dataset was divided into $4$ training, $2$ validation, and $4$ test patients. The number of glomeruli in each staining dataset was:  PAS - $662$ (train.), $588$ (valid.), $1092$ (test); Jones H\&E - $1043$ (test);H\&E - $1151$;  Sirius Red - $1049$ (test); CD34 - $1019$ (test); CD68 - $1046$ (test);CD3 - $1083$. The training set comprised all glomeruli from the source staining training patients ($662$) plus $4634$ tissue (i.e.\ non-glomeruli) patches (to account for the variance observed in non-glomeruli tissue).


\section{Results}
\label{sec:Results}
In the following, the proposed approach is referred to as Unsupervised Domain Augmentation using GANs (UDA-GAN), with two variants: with CycleGANs (UDA-CGAN) and StarGAN (UDA-\textasteriskcentered GAN). \color{black} These are presented and compared to the MDS1 \cite{gadermayrGenerativeAdversarialNetworks2019}, MDS2 \cite{gadermayrGenerativeAdversarialNetworks2019}, Unsupervised Domain Augmentation via Stain Decomposition (UDA-SD) \cite{lampert2019strategies} and vanilla PAS (vPAS), which is the direct application of a model trained on PAS to the target domains without any adaptation (i.e.\ the control).

\begin{figure}[ht]
	\begin{center}
	\small{
		\settoheight{\tempdima}{\includegraphics[width=0.105\textwidth]{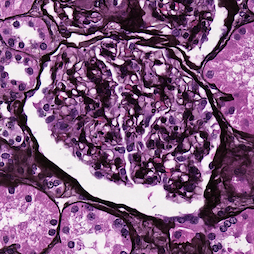}}%
		\begin{tabular}{ l@{ }c@{ }c@{ }c@{ }c@{ }
		}
			
			{ } & Original & \multicolumn{3}{c}{Translation to PAS}
			\\
			{ }&  & $1^{\textrm{st}}$ Epoch & $10^{\textrm{th}}$ Epoch & $20^{\textrm{th}}$ Epoch 
			\\
			\rowname{Jones H\&E } &
			
			\includegraphics[width=0.10\textwidth]{img/IFTA_Nx_0010_03_glomeruli_patch_59_orig.png} &
			\includegraphics[width=0.10\textwidth]{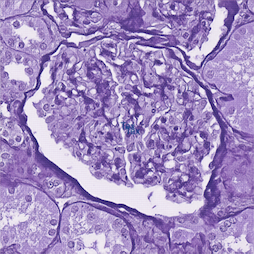} &
			\includegraphics[width=0.10\textwidth]{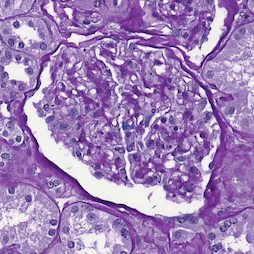} &
	    	\includegraphics[width=0.10\textwidth]{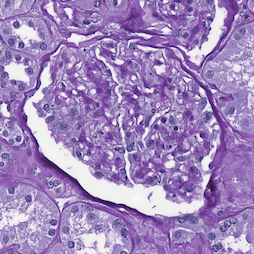}
			\\[-0.3ex]
			\rowname{CD68} &
			\includegraphics[width=0.10\textwidth]{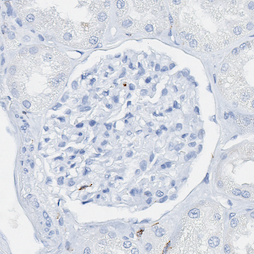} &
			\includegraphics[width=0.10\textwidth]{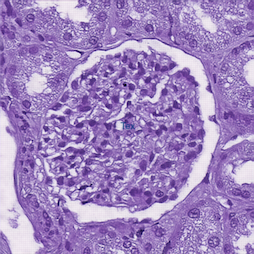} &
			\includegraphics[width=0.10\textwidth]{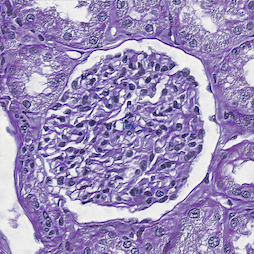} &
			\includegraphics[width=0.10\textwidth]{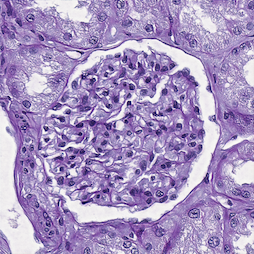}
			\\
		\end{tabular}
		}
		\caption{\color{black}Jones H\&E  and CD68 glomeruli patches translated to PAS using CycleGAN models from different training epochs.}
		\label{fig:cyclegan_epochs}
	\end{center}
\end{figure}

As mentioned, the CycleGAN models obtained after $50$ epochs are used, however, Brieu et al.\ \cite{Brieu2019DomainAA} show that realistic translations are obtained even in early epochs, which is confirmed in Figure \ref{fig:cyclegan_epochs}. They therefore train a segmentation model using translators taken from multiple epochs to increased augmentation variability. Therefore we include the Multi UDA-CGAN model in which translation models from each $5^\textrm{th}$ epoch are used, resulting in $40$ translation models.

For MDS1 and MDS2, the translated models are trained according to Section \ref{sec:StainTranslationModel}. \color{black} Variants of MDS1 and MDS2 using the StarGAN translators were evaluated, and are referred to as MDS\textasteriskcentered 1{} and MDS\textasteriskcentered 2. \color{black} The F$_1$-score, along with precision and recall, are used to measure the performance.
The presented results are the averages of five independent training repetitions, with corresponding standard deviations.

The experimental results are presented in Table \ref{tab: quantitative results}.
Baseline performances (U-Net models, as described above, trained and tested on the same stain using each stain's ground truth) were determined for each staining and presented in Table \ref{tab: baseline results}.

\begin{table*}[!ht]
	\centering
	\small{
	\begin{tabular}{L{2.5cm} L{1.2cm} C{1.7cm} C{1.7cm} C{1.7cm} C{1.7cm} C{1.7cm} C{1.7cm}}
		\hline \\[-2ex]
		\multirow{2}{*}{\shortstack{Training\\ Strategy}} & \multirow{2}{*}{Score} & \multicolumn{5}{c}{Test Staining}\\
		& & PAS & Jones H\&E & CD68 & Sirius Red & CD34 & Overall\\
		\hhline{========}
		\multirow{3}{*}{vPAS} & F$_1$ & \color{black}0.907 \footnotesize{(0.009)}\color{black} & 0.085 \footnotesize{(0.034)} & 0.001 \footnotesize{(0.002)} & 0.016 \footnotesize{(0.018)} & 0.071 \footnotesize{(0.063)} & 0.043 \footnotesize{(0.041)} \\
		& Precision & \color{black} 0.885 \footnotesize{(0.023)}\color{black}  & 0.055 \footnotesize{(0.021)} & 0.097 \footnotesize{(0.129)} & 0.034 \footnotesize{(0.034)} & 0.257 \footnotesize{(0.243)} & 0.111 \footnotesize{(0.101)}\\ 
		& Recall & \color{black} 0.932 \footnotesize{(0.014)}\color{black} & 0.418 \footnotesize{(0.316)} & 0.001 \footnotesize{(0.001)} & 0.073 \footnotesize{(0.101)} & 0.058 \footnotesize{(0.039)} & 0.137 \footnotesize{(0.190)}\\ 
		\hline
		\multirow{3}{1.9cm}{UDA-SD \cite{lampert2019strategies}} & F$_1$ & 0.891 \footnotesize{(0.007)} & 0.791 \footnotesize{(0.079)} & 0.147 \footnotesize{(0.048)} & 0.828 \footnotesize{(0.046)} & 0.739 \footnotesize{(0.026)} & 0.679 \footnotesize{(0.303)}\\
		& Precision & 0.840 \footnotesize{(0.018)} & 0.699 \footnotesize{(0.116)} & 0.365 \footnotesize{(0.238)} & 0.778 \footnotesize{(0.088)} & 0.695 \footnotesize{(0.054)} & 0.675 \footnotesize{(0.183)}\\ 
		& Recall & 0.950 \footnotesize{(0.007)} & 0.926 \footnotesize{(0.015)} & 0.099 \footnotesize{(0.032)} & 0.892 \footnotesize{(0.024)} & 0.795 \footnotesize{(0.059)} & 0.732 \footnotesize{(0.359)}\\ 
		\hline
		\multirow{3}{1.9cm}{MDS1\cite{gadermayrGenerativeAdversarialNetworks2019}} & F$_1$& - & \textbf{0.872} \footnotesize{(0.016)} & 0.395 \footnotesize{(0.057)} & 0.828 \footnotesize{(0.040)} & 0.673 \footnotesize{(0.033)} & 0.692 \footnotesize{(0.215)} \\
		& Precision & - & 0.843 \footnotesize{(0.036)} & 0.447 \footnotesize{(0.092)} &0.787 \footnotesize{(0.071)} &0.857 \footnotesize{(0.033)} & 0.734 \footnotesize{(0.193)}\\ 
		& Recall & - & 0.904 \footnotesize{(0.018)} & 0.364 \footnotesize{(0.071)} & 0.877 \footnotesize{(0.020)} &0.556 \footnotesize{(0.047)} &0.675 \footnotesize{(0.261)}\\ 
		\hline
		\multirow{3}{1.9cm}{MDS2\cite{gadermayrGenerativeAdversarialNetworks2019}} & F$_1$ & - & 0.869 \footnotesize{(0.020)}  & 0.586 \footnotesize{(0.059)} & 0.797 \footnotesize{(0.040)} & 0.739 \footnotesize{(0.044)} & 0.748 \footnotesize{(0.121)}\\
		& Precision & - & 0.833 \footnotesize{(0.049)} & 0.519 \footnotesize{(0.108)} &0.699 \footnotesize{(0.061)} &0.723 \footnotesize{(0.051)} &0.695 \footnotesize{(0.132)}\\ 
		& Recall & - & 0.909 \footnotesize{(0.013)} & 0.697 \footnotesize{(0.059)} & 0.929 \footnotesize{(0.004)} & 0.765 \footnotesize{(0.106)} &0.825 \footnotesize{(0.112)}\\ 
		\hline
		\multirow{3}{*}{\shortstack{UDA-CGAN}} & F$_1$ & \textbf{0.901} \footnotesize{(0.011)}  & 0.856 \footnotesize{(0.036)} & \textbf{0.705} \footnotesize{(0.031)} & \textbf{0.873} \footnotesize{(0.025)} & {0.799} \footnotesize{(0.034)} & \textbf{0.827} \footnotesize{(0.078)} \\
		& Precision & 0.869 \footnotesize{(0.034)} & 0.800 \footnotesize{(0.069)} & 0.690 \footnotesize{(0.059)} & 0.830 \footnotesize{(0.051)} & 0.754 \footnotesize{(0.076)}  & 0.789 \footnotesize{(0.069)}\\ 
		& Recall & 0.936 \footnotesize{(0.014)} & 0.924 \footnotesize{(0.012)} & 0.723 \footnotesize{(0.034)} & 0.922 \footnotesize{(0.009)} & 0.856 \footnotesize{(0.036)}  & 0.872 \footnotesize{(0.089)} \\  
		\hline
		\multirow{3}{\linewidth}{\shortstack{Multi UDA-CGAN}} & F$_1$ & 0.897 \footnotesize{(0.010)} & 0.863 \footnotesize{(0.030)} & 0.684 \footnotesize{(0.046)} & 0.861 \footnotesize{(0.021)} & \textbf{0.808} \footnotesize{(0.023)} & 0.822 \footnotesize{(0.084)} \\
		& Precision & 0.860 \footnotesize{(0.021)} &0.812 \footnotesize{(0.057)} & 0.648 \footnotesize{(0.098)} &0.813 \footnotesize{(0.043)} & 0.764 \footnotesize{(0.061)} & 0.779 \footnotesize{(0.081)}\\ 
		& Recall & 0.937 \footnotesize{(0.007)} & 0.922 \footnotesize{(0.009)} &0.736 \footnotesize{(0.038)} &0.917 \footnotesize{(0.010)} &0.862 \footnotesize{(0.032)} & 0.875 \footnotesize{(0.083)}\\ 
		\hline
		\multirow{3}{*}{MDS\textasteriskcentered 1} & F$_1$& - & 0.756 \footnotesize{(0.086)} & 0.092 \footnotesize{(0.055)} & 0.599 \footnotesize{(0.108)} & 0.751 \footnotesize{(0.033)} & 0.550 \footnotesize{(0.314)} \\
		& Precision & - & 0.675 \footnotesize{(0.136)} & 0.242 \footnotesize{(0.116)} &0.496 \footnotesize{(0.123)} &0.742 \footnotesize{(0.092)} & 0.539 \footnotesize{(0.223)}\\ 
		& Recall & - & 0.881 \footnotesize{(0.029)} & 0.061 \footnotesize{(0.044)} & 0.780 \footnotesize{(0.099)} & 0.774 \footnotesize{(0.070)} &0.624 \footnotesize{(0.379)}\\ 
		\hline\multirow{3}{*}{MDS\textasteriskcentered 2} & F$_1$ & - & 0.816 \footnotesize{(0.060)}  & 0.525 \footnotesize{(0.048)} & 0.837 \footnotesize{(0.032)} & 0.766 \footnotesize{(0.030)} & 0.736 \footnotesize{(0.144)}\\
		& Precision & - & 0.740 \footnotesize{(0.096)} & 0.874 \footnotesize{(0.037)} &0.785 \footnotesize{(0.059)} & 0.752 \footnotesize{(0.030)} & 0.787 \footnotesize{(0.061)}\\ 
		& Recall & - & 0.918 \footnotesize{(0.008)} & 0.376 \footnotesize{(0.046)} & 0.901 \footnotesize{(0.014)} & 0.785 \footnotesize{(0.073)} & 0.745 \footnotesize{(0.253)}\\ 
		\hline
		\multirow{3}{*}{\shortstack{UDA-\textasteriskcentered GAN}} & F$_1$ & 0.890 \footnotesize{(0.022)}  & 0.807 \footnotesize{(0.031)} & 0.549 \footnotesize{(0.081)} & 0.792 \footnotesize{(0.052)} & 0.758 \footnotesize{(0.076)} & 0.759 \footnotesize{(0.127)} \\
		& Precision & 0.853 \footnotesize{(0.043)} & 0.717 \footnotesize{(0.050)} & 0.794 \footnotesize{(0.044)} & 0.703 \footnotesize{(0.085)} & 0.738 \footnotesize{(0.082)}  & 0.761 \footnotesize{(0.062)}\\ 
		& Recall & 0.933 \footnotesize{(0.008)} & 0.926 \footnotesize{(0.010)} & 0.426 \footnotesize{(0.090)} & 0.913 \footnotesize{(0.013)} & 0.796 \footnotesize{(0.135)}  & 0.799 \footnotesize{(0.216)} \\ 
		\hline
	\end{tabular}
	}
	\caption{Quantitative results for each strategy trained on PAS (source staining) and tested on different (target) stainings. Standard deviations are in parentheses, the highest F$_1$ scores for each staining are in bold (\color{black}PAS is not included in the vPAS average since it is the training staining.) \color{black}}
	\label{tab: quantitative results}
\end{table*}

\begin{table*}[!ht]
 \centering
 \small{
 \begin{tabular}{L{1.7cm} C{1.65cm} C{1.65cm} C{1.65cm} C{1.65cm} C{1.65cm} C{1.65cm} C{1.5cm}}
  \hline \\[-2ex]
  & PAS & Jones H\&E & CD68 & Sirius Red & CD34 & Overall\\
  \hhline{========}
   F$_1$ & 0.907 \footnotesize{(0.009)} & 0.864 \footnotesize{(0.011)} & 0.853 \footnotesize{(0.018)} & 0.867 \footnotesize{(0.016)} & 0.888 \footnotesize{(0.015)} & 0.876 \footnotesize{(0.022)} \\
                       Precision & 0.885  \footnotesize{(0.023)} & 0.824   \footnotesize{(0.020)} &  0.846 \footnotesize{(0.027)} & 0.801  \footnotesize{(0.042)} & 0.862 \footnotesize{(0.015)} & 0.844  \footnotesize{(0.032)}\\ 
                       Recall & 0.932  \footnotesize{(0.014)} & 0.911 \footnotesize{(0.005)} & 0.856 \footnotesize{(0.022)} & 0.957 \footnotesize{(0.018)} & 0.929 \footnotesize{(0.011)} &  0.917 \footnotesize{(0.038)}\\
  \hline
 \end{tabular}
 }
  \caption{Quantitative baseline results (standard deviations are in parentheses).}
  \label{tab: baseline results}
\end{table*}

 Despite the fact that the translations obtained using both CycleGAN and StarGAN look realistic (see Figure \ref{fig:cyclegan_translation}), it can be observed that the direction of translation (MDS1 vs MDS2) and translation model (StarGAN vs CycleGAN) influence the results. This is best illustrated with MDS1, in which a model trained on the original PAS data is applied to target data translated to PAS. It can be observed that in each target stain, the difference between CycleGAN and StarGAN translations is significant, although there appears to be no significant difference in the quality of the translations. The proposed UDA-GAN approaches show more stable performance while the best results are obtained using UDA-CGAN.

Furthermore, UDA-CGAN reaches almost baseline performance in three out of five test stainings. Despite the fact that the model has seen data from PAS stain only $20\%$ of the time during training, the model has baseline performance on this (source) domain.
The model also approaches baseline performance in target stains Jones and Sirius Red. For stains CD68 and CD34, the model reaches an F$_1$-score of $0.705$ and $0.799$, meaning that it gives an improvement of $11.9\%$ and $ 6\%$ respectively over the next best CycleGAN method (MDS2). The average performance over the five different stainings show that UDA-CGAN reaches an average F$_1$-score of $0.827$ ($0.808 $ without including the PAS staining, in order to be fairly compared to the MDS approaches), while MDS2, as the next best method, reaches an F$_1$-score of $0.748$. The biggest relative difference is observed in staining CD68 where the overall improvement is $55.8\%$ compared to the original approach \cite{lampert2019strategies} and $11.9\% $ compared to MDS2. Other than the baseline, UDA-CGAN is the only to achieve acceptable results in this staining.
Multi UDA-CGAN does not improve upon this, possibly because it introduces too much variability. 

\subsection{Unseen Stains}
\label{sec:unseen}

\begin{figure}[ht]
	\begin{center}
		\settoheight{\tempdima}{\includegraphics[width=1.6cm]{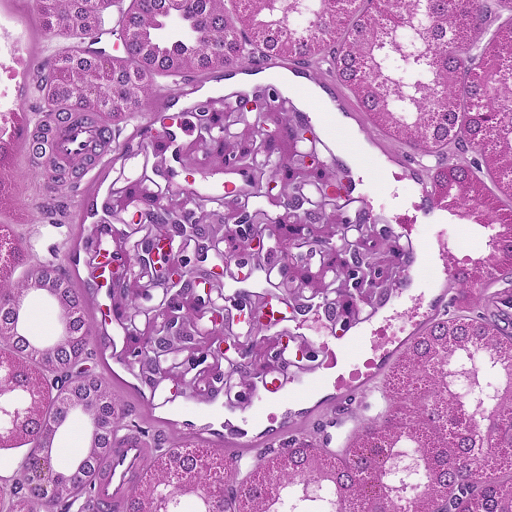}}%
		\begin{tabular}{c@{ }c@{ }}
			H\&E & CD3
			\\
			\includegraphics[width=0.10\textwidth]{img/IFTA_16_01_glomeruli_patch_104.png} &
			\includegraphics[width=0.10\textwidth]{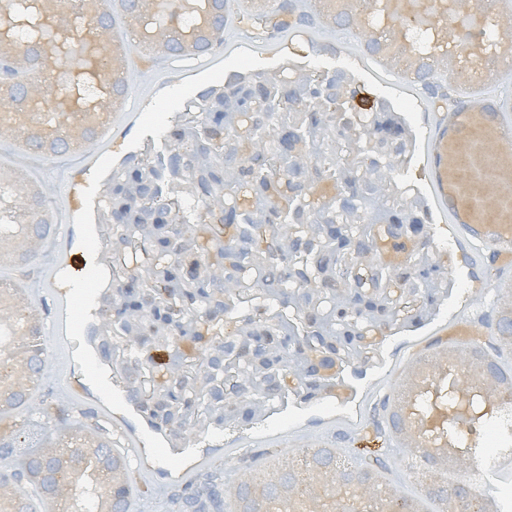}
		\end{tabular}
		\caption{\color{black}Examples of glomeruli from the unseen stains.}
		\label{fig:unseen_stains}
	\end{center}
\end{figure}

In order to further evaluate the stain invariance of the UDA-GAN approaches, they are tested on two unseen stains (unseen to both the CycleGAN/StarGAN and UDA-GAN models), see Figure \ref{fig:unseen_stains}: histological stain H\&E (a general overview staining not specific for a protein) and immunohistochemical stain CD3 (T cell marker). Even though they highlight similar structures to the `virtually' seen stains, they are visually very different in appearance. For each stain, images are taken from $3$ patients containing $1151$ (H\&E) and $1083$ (CD3) glomeruli. 
Table \ref{tab:unseen_results} presents the results, averaged over the previously trained UDA-GAN models.

\begin{table}[!ht]
	\centering
	\small{
		\begin{tabular}{L{1.5cm} L{1.2cm} C{1.2cm} C{1.2cm} C{1.2cm}}
			\hline \\[-2ex]
			& & H\&E & CD3 & Average\\
			\hhline{=====}
			\multirow{3}{*}{vPAS} & F$_1$ & 0.126 \footnotesize{(0.066)} & 0.000 \footnotesize{(0.000)}  &  0.063 
			\\
			& Precision & 0.070  \footnotesize{(0.041)} & 0.018  \footnotesize{(0.018)}   & 0.044   
			\\ 
			& Recall &  0.854  \footnotesize{(0.109)} &  0.000 \footnotesize{(0.000)}  &  0.427 
			\\
			\hline\\[-2ex]
			\multirow{3}{*}{\shortstack{UDA-CGAN}}& F$_1$ & 0.731  \footnotesize{(0.100)} & 0.658 \footnotesize{(0.075)} & 0.694 
			\\
			& Precision & 0.781 \footnotesize{(0.122)}  & 0.569   \footnotesize{(0.124)} &  0.675 
			\\ 
			& Recall & 0.697 \footnotesize{(0.123)} & 0.802 \footnotesize{(0.042)} & 0.750 
			\\
			\hline\\[-2ex]
			\multirow{3}{*}{\shortstack{UDA-\textasteriskcentered GAN}} & F$_1$ & 0.752 \footnotesize{(0.087)} & 0.650 \footnotesize{(0.030)}  & 0.701 
			\\
			& Precision &  0.824 \footnotesize{(0.057)} & 0.853 \footnotesize{(0.065)}   & 0.839 
			\\ 
			& Recall &  0.706 \footnotesize{(0.147)} & 0.531 \footnotesize{(0.058)}  & 0.618 
			\\
			
			\hline
		\end{tabular}
	}
	\caption{Quantitative results of UDA-GAN models on unseen stains (standard deviations are in parentheses).}
	\label{tab:unseen_results}
\end{table}

Although the results are lower than those obtained using stains virtually seen during training, i.e.\ stain translation targets during augmentation, they confirm the network's capacity for stain invariant segmentation.

On average, both UDA-CGAN and UDA-\textasteriskcentered GAN perform equally well on unseen stains. And when taken in context of Table \ref{tab: quantitative results}, although they generally achieve lower results, UDA-\textasteriskcentered GAN remains within the range previously seen and UDA-CGAN exhibits more variance.

A high $F_1$ score is achieved when faced with a completely new stain colour profile (H\&E). When faced with a similar stain profile to one virtually seen (CD3, similar to CD68), the corresponding UDA-CGAN $F_1$ score is similar and UDA-\textasteriskcentered GAN improves, likely because CD3 has more contrast (due to an unspecific reaction of the primary antibody) when compared to CD68, with which it struggles. It is worth noting that the models with the best PAS performance (which can be determined since annotations exist) are also those that perform best on the unseen stains.

\section{Further Analysis and Discussion}
\label{sec:Discussion}
In this section the results are analysed in more detail, including calculating additional statistics and, and related to those found in the literature. Further analysis of the trained models is presented by visualising their feature distributions and their attention. Finally, the topic of CycleGAN training is discussed.

\subsection{Translation Direction}

\label{sec:discussion_stain_invariant}
Gadermayr et al.\ suggest that translation ``should always be performed from the difficult-to-segment to the easy-to-segment domain'' and that MDS1 is the preferred method \cite{gadermayrGenerativeAdversarialNetworks2019}. Table \ref{tab: baseline results} shows that the difficult-to-segment stain is CD68 and the easy is PAS as it results in a more accurate baseline segmentation. It is also observed in Table \ref{tab: quantitative results} that MDS1 with PAS as the source domain is in fact surpassed by MDS2 in all but one target stainings (in which it is equal).
These results suggest that the characteristics of ``difficult-to-segment'' stainings may vary, and the method/translation direction should be adjusted to the specific requirements of a given biological question, i.e.\ the panel of necessary staining methods.

\subsection{Representation Comparison}
\label{subsec:repcomp}
In the case of staining CD68, neither MDS1, \color{black} MDS\textasteriskcentered 1, MDS2, nor, MDS\textasteriskcentered 2 \color{black} perform well. Poor MDS1 \color{black} (MDS\textasteriskcentered 1) \color{black} performance could indicate that the CycleGAN \color{black} (StarGAN) \color{black} translation between the CD68 and PAS domains do not capture the features the PAS model uses for segmentation. On the other hand, poor MDS2 \color{black} and MDS\textasteriskcentered 2 \color{black} performance could indicate that the translation between PAS and CD68 contains features that are not present in real CD68. From a biological viewpoint, this most likely represents the fact that immunohistochemistry for CD68 highlights just one specific, migratory cell population (macrophages) that is not part of the pre-existing tissue architecture, with a brown chromogen, while the anatomical structures are only faintly stained (blue ``counterstain'' using hemalaun). Strikingly, immunohistochemistry for CD34, a marker for vascular endothelial cells, labeled with a red chromogen, performs much better. This can probably be explained by a specific immunohistochemical labeling of anatomical structures (blood vessels) in addition to the blueish counterstain, containing more features that are also covered in the other staining methods (PAS, Jones H\&E). \color{black} This is particularly evident with StarGAN, which exploits common inter-stain characteristics. 

Between MDS1 and MDS2 (both CycleGAN and StarGAN translations) the largest difference is seen in stain CD68, which marks a protein exclusively produced by macrophages. PAS, as a chemical reaction staining glycolysated proteins in general, highlights a part of macrophages (co-located, but not overlapping, with CD68). Thus the translation from PAS to CD68 (MDS2, MDS\textasteriskcentered2) is easier than the reverse (MDS1, MDS\textasteriskcentered1), since PAS contains some (but not all) of the information exposed by CD68.
\color{black}

The fact that UDA-CGAN outperforms both MDS1 and MDS2 using the same translation functions indicates that it is capable of extracting more general (stain invariant) features, i.e.\ it avoids learning stain specific and `false' features introduced by the CycleGAN. \color{black} This is also the case for UDA-\textasteriskcentered GAN but to a lesser extent. UDA-\textasteriskcentered GAN uses the same generator for all translations so it is likely to extract similar features between the source stain and all target stains. This reduces the impact of the multi-stain augmentation, which becomes evident when comparing UDA-\textasteriskcentered GAN to MDS\textasteriskcentered2. 

\subsubsection{Feature Distributions}
\begin{figure*}[t]
	\begin{center}
		\settoheight{\tempdima}{\includegraphics[width=0.3\textwidth]{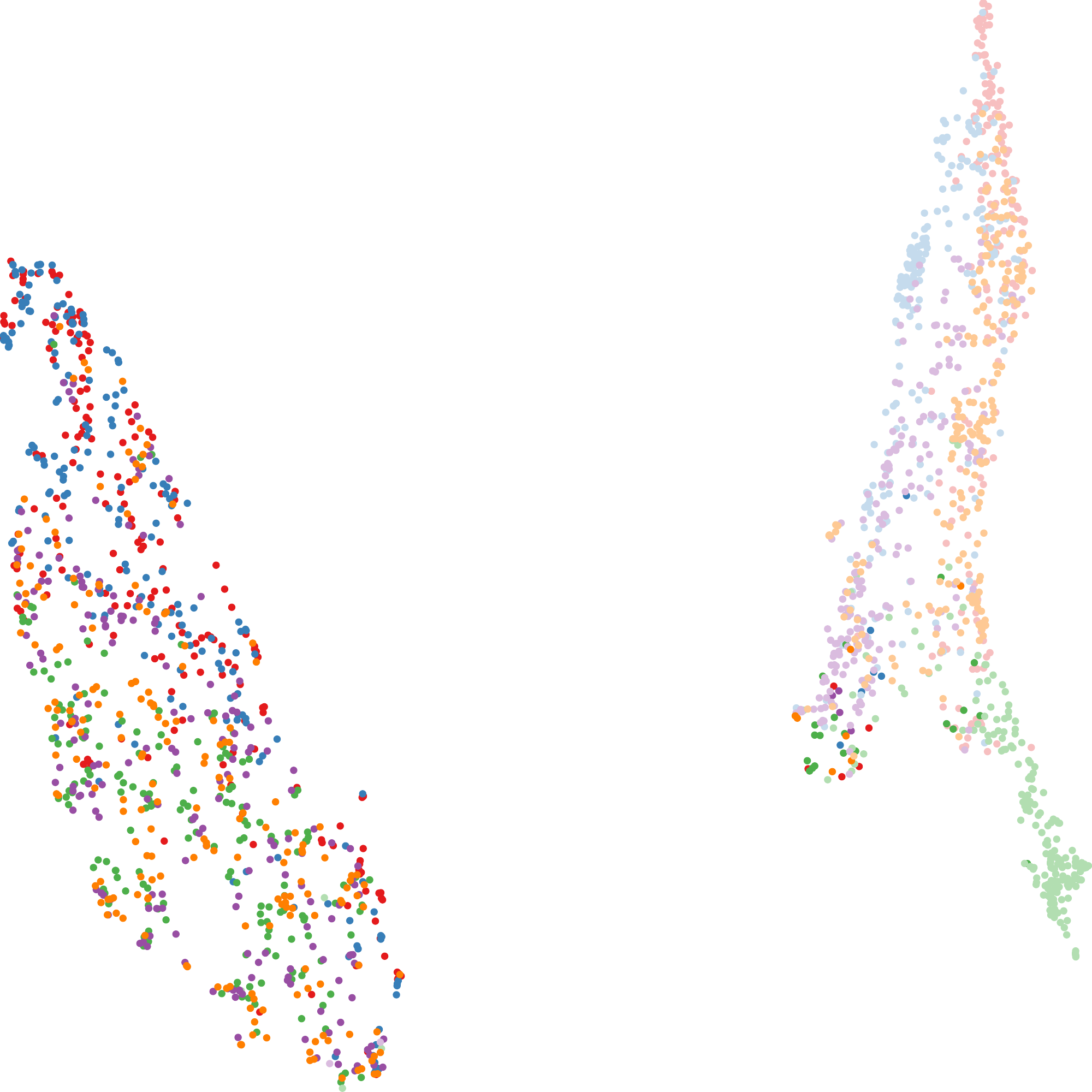}}
		\begin{tabular}{c@{ }c@{ }c@{ }c@{ }}
			UDA-CGAN & UDA-\textasteriskcentered GAN & MDS1 (best PAS model) & MDS\textasteriskcentered 1 (best PAS model)\\
			\includegraphics[width=0.24\textwidth,frame]{img/umap/gan_clean_features_activation_22_pas_test_data.pdf} &
			\includegraphics[width=0.24\textwidth,frame]{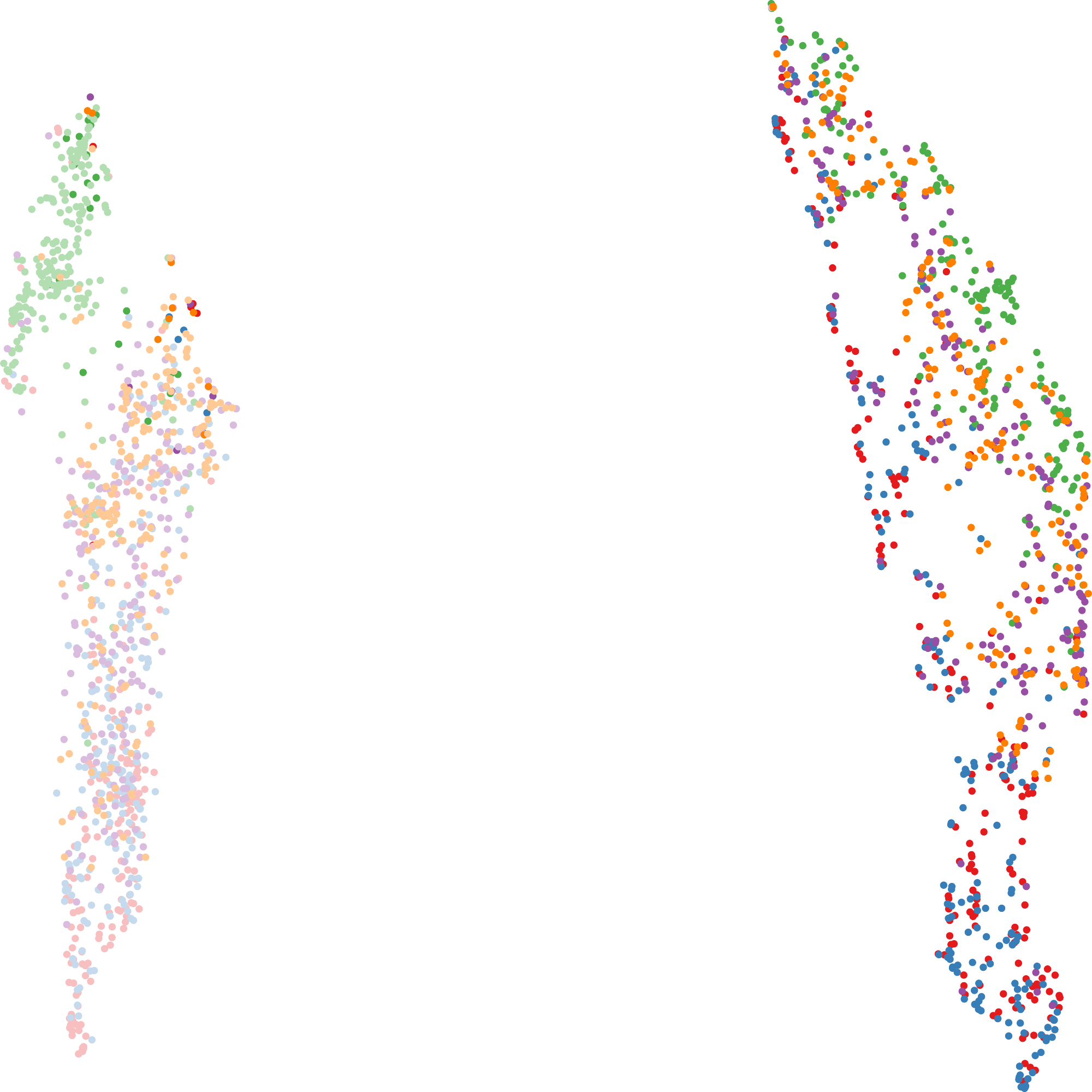} &
			\includegraphics[width=0.24\textwidth,frame]{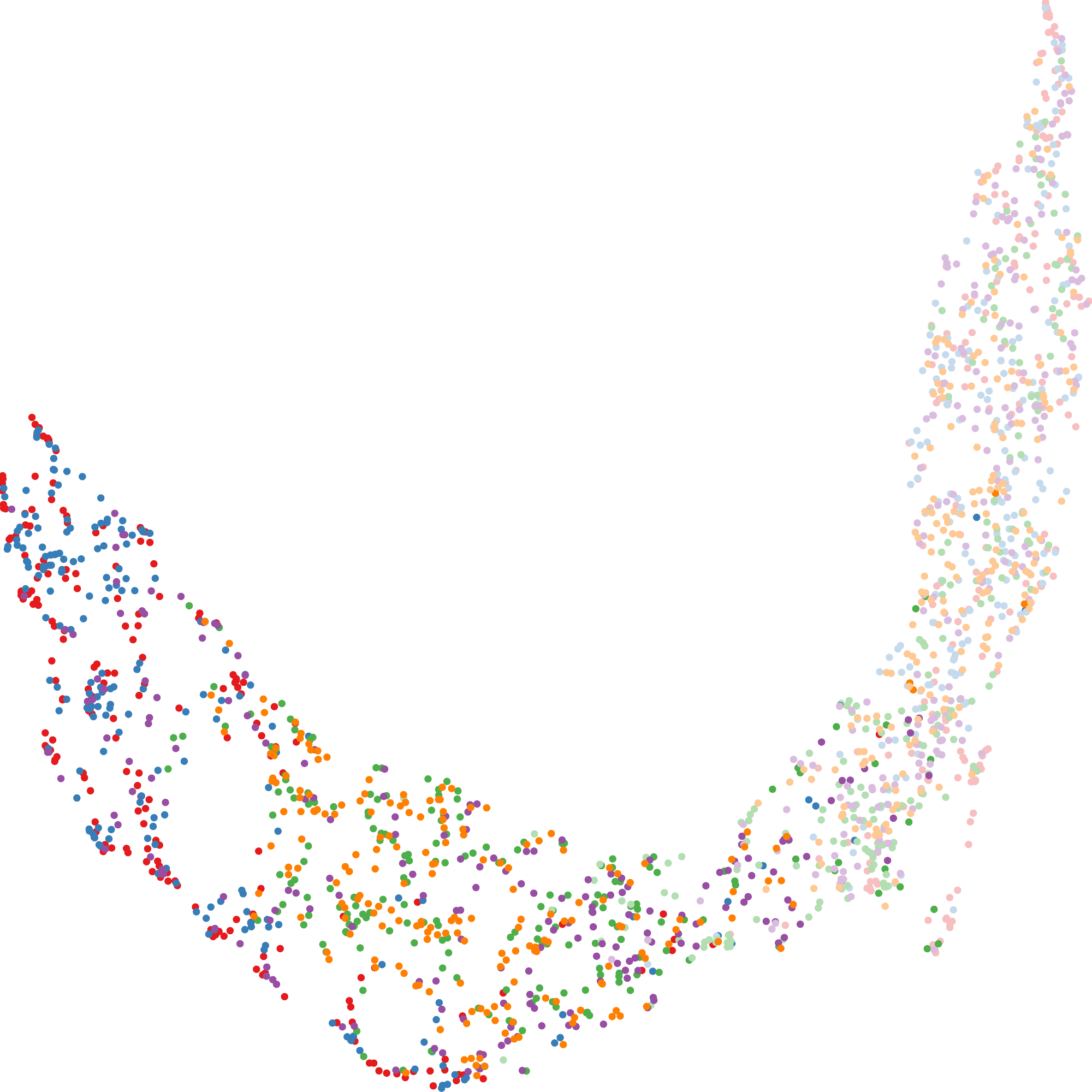} &
			\includegraphics[width=0.24\textwidth,frame]{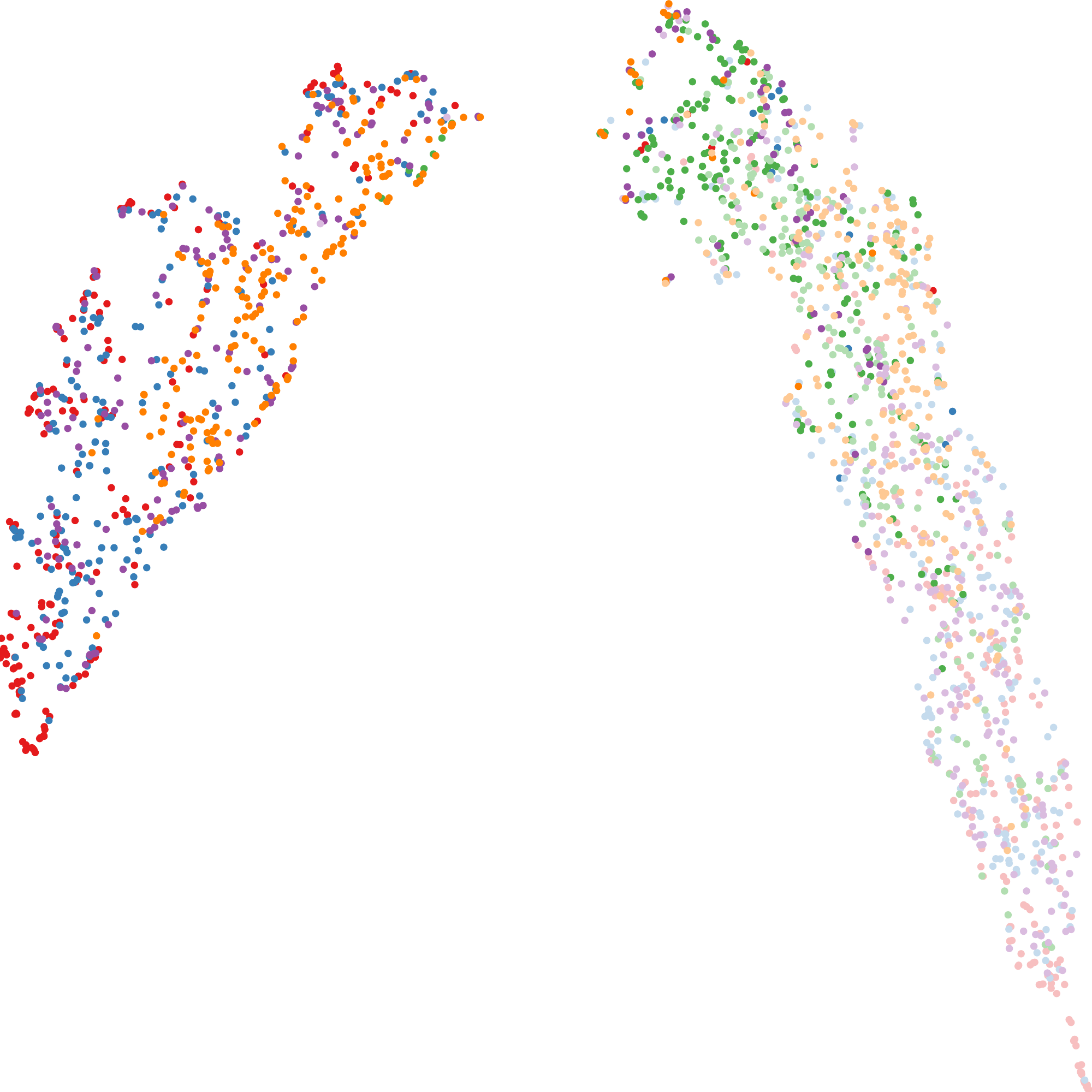} \\
			\multicolumn{4}{c}{\includegraphics[width=0.66\textwidth]{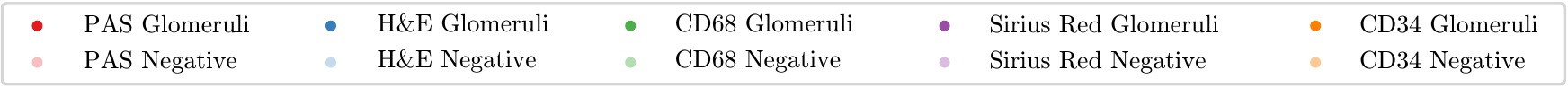}}
		\end{tabular}
		\caption{Two-dimensional UMAP embeddings of the representation learnt, sampled from the penultimate convolutional layer using $200$ patches per stain per class from the overall best performing PAS and UDA-CGAN models. Each point represents a patch from the respective class and staining (glomeruli patches are centred on a glomeruli).}
		\label{fig:tsne_rgb_vs_gan}
	\end{center}
\end{figure*}
In order to visualise the distributions of each model's extracted features, Figure \ref{fig:tsne_rgb_vs_gan} presents UMAP (Uniform Manifold Approximation and Projection) embeddings \cite{Umap2018} of the penultimate convolutional layer's activations in the best performing model (over all stainings) for two hundred random glomeruli and two hundred random tissue patches of each staining using MDS1, MDS\textasteriskcentered 1, UDA-CGAN, and UDA-\textasteriskcentered GAN (MDS2 and MDS\textasteriskcentered 2 models are stain specific, therefore cannot be applied to all stainings). For MDS1 and MDS\textasteriskcentered 1, the translations to the PAS stain is achieved using CycleGAN and StarGAN respectively, and UDA-CGAN and UDA-\textasteriskcentered GAN are applied to original data without any modification. \color{black}

\begin{table*}[t]
	\centering
\small{
	\begin{tabular}{L{1.8cm} L{2.3cm} C{1.8cm} C{1.8cm} C{1.8cm} C{1.8cm} C{1.8cm} }
		\hline \\[-2ex]
		 & Training Strategy	& PAS & Jones H\&E & CD68 & Sirius Red & CD34 \\
		\hhline{=======}
		\multirow{4}{\linewidth}{Glomeruli-Combined} & UDA-CGAN & 0.069 \footnotesize{(0.004)}&{0.106} \footnotesize{(0.008)}&\textbf{0.000} \footnotesize{(0.007)}&-0.044 \footnotesize{(0.003)}&\textbf{-0.019} \footnotesize{(0.003)}
		\\
		& UDA-\textasteriskcentered GAN & 
		\textbf{0.051} \footnotesize{(0.008)}&
		\textbf{0.094} \footnotesize{(0.005)}&
		0.069 \footnotesize{(0.010)}&
		-0.041 \footnotesize{(0.003)}&
		-0.024 \footnotesize{(0.003)}\\
		& MDS1 & 0.198 \footnotesize{(0.011)} & 0.176 \footnotesize{(0.014)} & -0.057 \footnotesize{(0.002)} & {-0.038} \footnotesize{(0.004)} & -0.070 \footnotesize{(0.005)}
		\\
		& MDS\textasteriskcentered 1 & 
		0.219 \footnotesize{(0.006)}&
		0.048 \footnotesize{(0.011)}&
		0.166 \footnotesize{(0.004)}&
		\textbf{-0.027} \footnotesize{(0.003)}&
		-0.063 \footnotesize{(0.004)}\\
		\hline
		\multirow{4}{\linewidth}{Glomeruli-{PAS Glomeruli}} & UDA-CGAN  & - & 0.004 \footnotesize{(0.002)}&\textbf{0.123} \footnotesize{(0.008)} &{0.078} \footnotesize{(0.003)}&\textbf{0.071} \footnotesize{(0.006)}
		\\
		& UDA-\textasteriskcentered GAN & -
		&
		0.004 \footnotesize{(0.001)}&
		0.179 \footnotesize{(0.003)}&
		\textbf{0.070} \footnotesize{(0.002)}&
		0.090 \footnotesize{(0.003)}
		\\
		& MDS1 & - &\textbf{0.002} \footnotesize{(0.001)} &0.253 \footnotesize{(0.007)}&0.175 \footnotesize{(0.009)} &0.255 \footnotesize{(0.003)}
		\\ 
		& MDS\textasteriskcentered 1 & -
		&
		0.037 \footnotesize{(0.005)}&
		0.477 \footnotesize{(0.007)}&
		0.098 \footnotesize{(0.003)}&
		0.186 \footnotesize{(0.010)}
		\\
		\hhline{=======}
		\color{black}
		\multirow{4}{\linewidth}{Glomeruli-Tissue} & UDA-CGAN  & 0.594 \footnotesize{(0.009)} & 0.567 \footnotesize{(0.017)} & \textbf{0.533} \footnotesize{(0.009)} & 0.554 \footnotesize{(0.012)} & \textbf{0.551} \footnotesize{(0.010)}
		\\
		& UDA-\textasteriskcentered GAN & 
		\textbf{0.625} \footnotesize{(0.016)} &
		\textbf{0.584} \footnotesize{(0.015)} &
		0.475 \footnotesize{(0.014)} &
		\textbf{0.581} \footnotesize{(0.012)} &
		0.530 \footnotesize{(0.005)}
		\\
		& MDS1 & 0.489 \footnotesize{(0.007)} & 0.481 \footnotesize{(0.017)} &0.300 \footnotesize{(0.009)} & 0.300 \footnotesize{(0.016)} & 0.424 \footnotesize{(0.006)}
		\\ 
		& MDS\textasteriskcentered 1 & 
		0.489 \footnotesize{(0.007)} &
		0.456 \footnotesize{(0.019)} &
		{0.051} \footnotesize{(0.004)} &
		0.354 \footnotesize{(0.023)} &
		0.449 \footnotesize{(0.005)}
		\\
		\hline
	\end{tabular}
	}
	\caption{Silhouette scores measuring (averaged over 3 different random samplings): the separation between the each stain's glomeruli class and the glomeruli class formed from all other stains (Combined), the separation between the each stain's glomeruli class and the PAS glomeruli class (PAS), and the separation between each stain's glomeruli class and its negative class. Calculated using the features of 200 random patches per stain per class, extracted from the penultimate convolutional layer: 0 means total overlap, 1 means total separation, -1 means that samples are more similar to the other class than their own, therefore values closer to 0 are better (in bold) for Glomeruli-Combined and Glomeruli-PAS Glomeruli, and 1 is better (in bold) for Glomeruli-Tissue.}
	\label{tab: silhuette_score}
\end{table*}

\color{black} In order to quantitatively measure these distributions, silhouette scores \cite{ROUSSEEUW198753} have been calculated between:
\begin{itemize}
    \item each stain's glomeruli class and the union of the glomeruli samples from all other stains;
    \item PAS glomeruli and each target stain's glomeruli;
    \item each stain's glomeruli and negative samples.
\end{itemize}
These are presented in Table \ref{tab: silhuette_score}. The first should favour the UDA-GAN approaches since their goal is to learn a stain invariant representation, the second should favour MDS1 and MDS\textasteriskcentered1, since their objective is to translate the target stainings to the PAS distribution, while the third should be the goal of all approaches.

In the first case, the UDA-GAN approaches exhibit larger (or equal) overlap between the glomeruli in all stainings, indicating greater clustering, which is reflected in the fact that these models are able to segment all stainings. Higher scores for the MDS1 approaches indicate less concentrated clustering (e.g.\ Jones H\&E , which appears to be concentrated in one part of the glomeruli space, separate from the other stains). 
 
Since the MDS1 approaches are trained on PAS, they rely on accurate translation models, which must result in a direct overlap with the PAS distribution. Whereas the UDA-GAN approaches can tolerate more translation variance (that does not result in glomeruli - tissue overlap) since they are trained on the translated data. 
Interestingly the scores for MDS1 and MDS\textasteriskcentered1 for PAS are relatively high, indicating that this approach fails to completely overlap the target stains with PAS.

In the second case, it can be observed that UDA-GAN approaches have better (or equal) overlap with the PAS glomeruli in all stainings. The score for MDS\textasteriskcentered1 in CD68 is much higher then for any other stain, and Figure \ref{fig:tsne_rgb_vs_gan} shows that the CD68 glomeruli class has been merged with the negative class, explaining the very low recall in Table \ref{tab: quantitative results}. 

In the third case, it can be observed that the UDA-GAN approaches better separate the glomeruli and tissue classes. This is especially illustrated in the case of CD68 in which UDA-\textasteriskcentered GAN learns to separate the glomeruli and tissue classes, whereas MDS\textasteriskcentered1 fails. As mentioned, MDS\textasteriskcentered1 relies on accurate translations, whereas the UDA-GAN approaches are able to correct for weak translations during training.

\subsubsection{Attention}

\begin{figure*}[ht]
	\begin{center}
		\settoheight{\tempdima}{\includegraphics[width=0.10\textwidth]{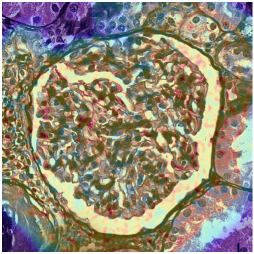}}%
		\begin{tabular}{@{}c@{ }c@{ }c@{ }c@{ }c@{ }c@{ }c@{ }c@{ }c@{ }c@{ }}
			&UDA-CGAN &UDA-\textasteriskcentered GAN & MDS1 & MDS\textasteriskcentered 1 & MDS2 & MDS\textasteriskcentered 2 & Baseline &
			\\
			\rowname{PAS}&
			\includegraphics[width=0.105\textwidth]{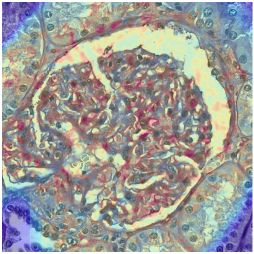} &
			\includegraphics[width=0.105\textwidth]{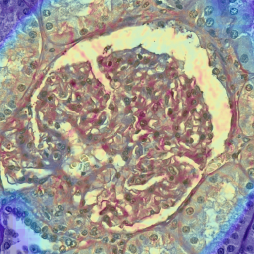} &
			 & 
			 &
			 &
			 &
			\includegraphics[width=0.105\textwidth]{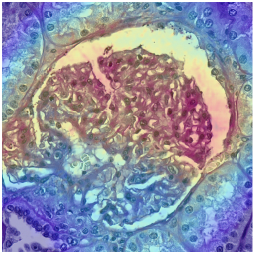} &
			\\
			\rowname{Jones H\&E } &
			\includegraphics[width=0.105\textwidth]{img/gradcam/IFTA_Nx_0010_03_glomeruli_patch_80_cl_2_gan.png}&
			\includegraphics[width=0.105\textwidth]{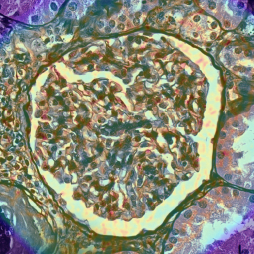}&
			\includegraphics[width=0.105\textwidth]{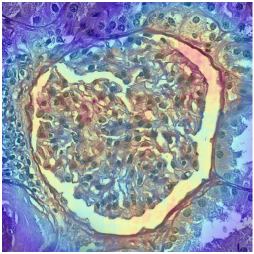}&
			\includegraphics[width=0.105\textwidth]{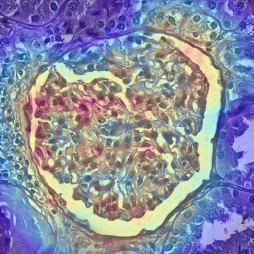}&
			\includegraphics[width=0.105\textwidth]{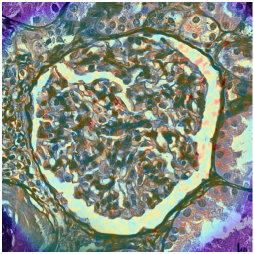}&
			\includegraphics[width=0.105\textwidth]{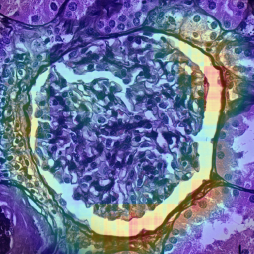}&
			\includegraphics[width=0.105\textwidth]{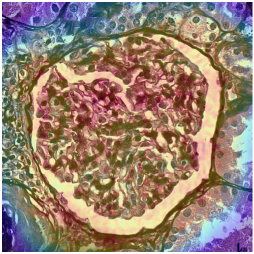}
			\\
			\rowname{CD68} &
			\includegraphics[width=0.105\textwidth]{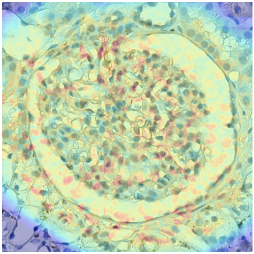}&
			\includegraphics[width=0.105\textwidth]{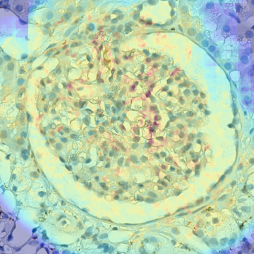}&
			\includegraphics[width=0.105\textwidth]{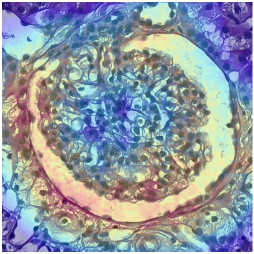}&
			 \includegraphics[width=0.105\textwidth]{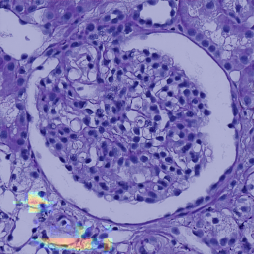}&
			\includegraphics[width=0.105\textwidth]{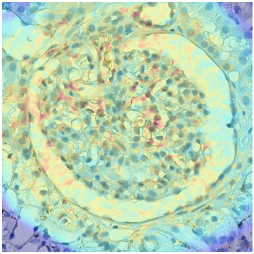}&
			\includegraphics[width=0.105\textwidth]{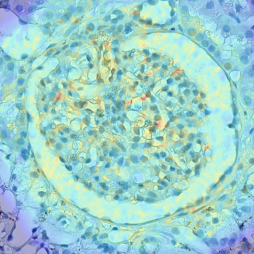}&
			\includegraphics[width=0.105\textwidth]{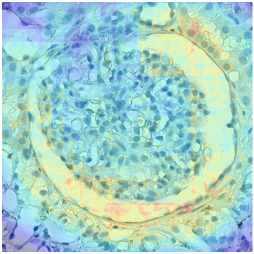}
			\\
			\rowname{Sirius Red} &
			\includegraphics[width=0.105\textwidth]{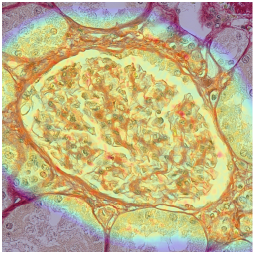}&
			\includegraphics[width=0.105\textwidth]{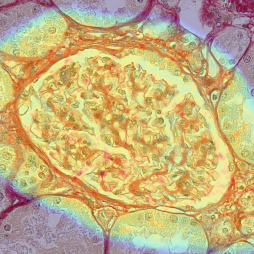}&
			\includegraphics[width=0.105\textwidth]{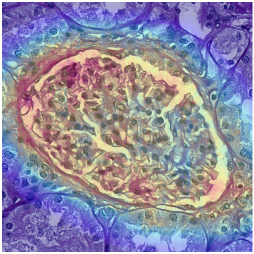}&
			\includegraphics[width=0.105\textwidth]{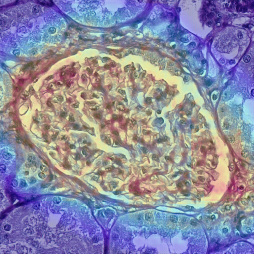}&
			\includegraphics[width=0.105\textwidth]{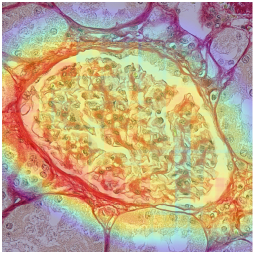}&
			\includegraphics[width=0.105\textwidth]{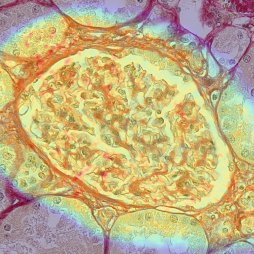}&
			\includegraphics[width=0.105\textwidth]{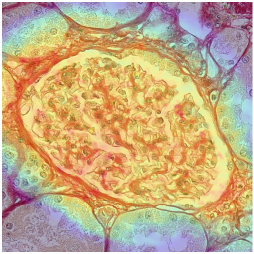}
   			\\
   			\rowname{CD34} &
   			\includegraphics[width=0.105\textwidth]{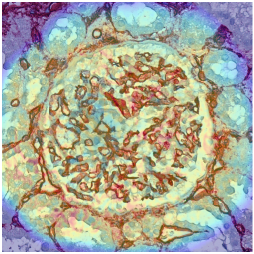}&
			 \includegraphics[width=0.105\textwidth]{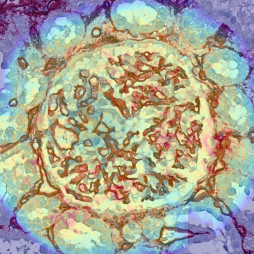} &
			 \includegraphics[width=0.105\textwidth]{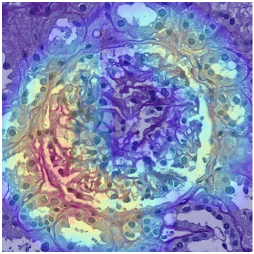} &
			 \includegraphics[width=0.105\textwidth]{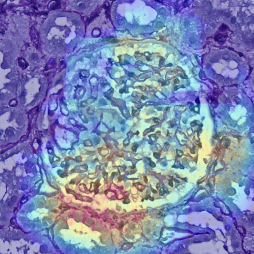} &
			 \includegraphics[width=0.105\textwidth]{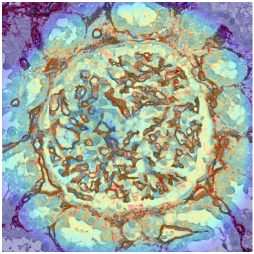} &
			 \includegraphics[width=0.105\textwidth]{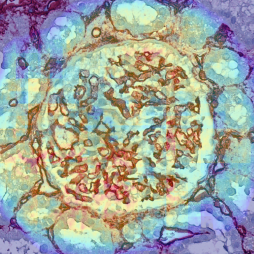} &
			 \includegraphics[width=0.105\textwidth]{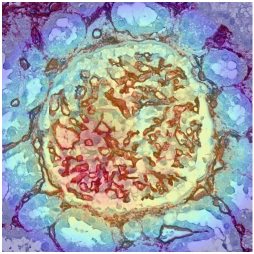}
			 
		\end{tabular}
		\caption{GradCAM visualisation of (column-wise) 1) overall best performing UDA-CGAN model; 2) best MDS1 model in each particular stain; 2) best MDS2 model in each particular stain; 3) best baseline model in each particular stain. N.B. the first column represents the attention of the same model over different stainings.}
		\label{fig:grad_cam_visualisation}
	\end{center}
\end{figure*}

Another approach to illustrate the representations learned is to use attention visualisations. Grad-CAM \cite{Selvaraju2017GradCAMVE} is used to visualise the attention of the penultimate convolutional layer, see Figure \ref{fig:grad_cam_visualisation}. For fair comparison, the best performing MDS1, MDS\textasteriskcentered 1, MDS2, and MDS\textasteriskcentered 2 model for each staining, and the baseline approaches are used. For UDA-CGAN and UDA-\textasteriskcentered GAN, the overall best-performing model is taken. 

The attention of the best-performing MDS1 (and MDS\textasteriskcentered 1, except CD68 for which  MDS\textasteriskcentered 1 does not work) models in all stainings is focused on border-like features, which the model uses in the original PAS domain (on which it was trained). Poor MDS1 and MDS\textasteriskcentered 1 performance can be explained by an absence of the specific features on which the trained model is focused on, and which are not necessarily present in the target domains nor relevant for the detection of the structures in general. When comparing attention in all the baseline models, it can be observed that in each staining, the models have a tendency to focus on stain specific features.
The attention of the MDS2 models is more general and close to the stain-invariant model's attention. According to the results presented in Table \ref{tab: quantitative results}, the stain-invariant approach (UDA-CGAN) gives an improvement in precision, while the recalls of both models are similar (in all stainings except CD34). Meaning that the stain-invariant model reduces false positives and, to some extent, false negatives (while both models detect true positives similarly). Thus, attention in the true positive class is expected to be similar. 
Both the MDS\textasteriskcentered 2 and UDA-\textasteriskcentered GAN use a common translation model (StarGAN), which, as previously mentioned, is biased towards common features between the stains. Thus both models are likely to focus on these common features, explaining their similar attentions (this is also confirmed in Table \ref{tab: quantitative results}). In addition to common features, the MDS2 models can also use stain-specific features because they are only exposed to one type of image (as is the case with the presented Jones H\&E example).

The main advantage of the UDA-GAN approaches is their ability to properly generalise across different stainings, as the presented attentions represent the features learnt by one model, while in all other cases a domain (stain) dependent model is obtained.

\subsubsection{Multi vs Single Stain Translation}

The fact that StarGAN has a single translator may force it to preserve common features between stains, features that UDA-\textasteriskcentered GAN is likely to focus on. It is also likely that these features are general and present in unseen stains, therefore UDA-\textasteriskcentered GAN's performance is similar within the virtually seen and unseen stains. Nevertheless, the single translator may force StarGAN to ignore hard-to-translate stains, e.g.\ CD68, and to minimise its loss upon the other stains. This would cause the general features to be extracted from the remaining stains. Low recall for CD68 and CD3 indicates that the model struggles to identify glomeruli, i.e.\ the general features do not exist in these stains. This is confirmed in Figure \ref{fig:tsne_rgb_vs_gan}, in which the CD68 glomeruli class overlaps the negative class. 
UDA-CGAN, on the other hand, is trained using the translators and has more sources of variation since each augmentation translator is independent. It can therefore leverage the additional information present in the augmented translations to achieve higher accuracy during application to them, while still learning a stain invariant representation. 

\subsection{Translation Quality}

\subsubsection{Quality Measure}

One way to validate the quality of such translations is to consult a pathologist, however this is expensive and time consuming. 
Preliminary validation by a pathologist confirms that the translations are indistinguishable from real images.

One of the most widely used measures for validating the quality of generated natural images is the Fre\'chet Inception Distance (FID) \cite{FID_Score}. 

FID measures the distance between features of the generated and real images extracted from the Inception network. Since this network is trained on natural images, it is not directly applicable to the medical domain. Instead, a new network could be trained on histpathological data and used as a feature extractor. However, due to (relatively) limited available data (in comparison to the ImageNet database which contains $\sim$14 million training images), it is likely that the trained network will capture dataset specific features (as confirmed by the experiments presented in this discussion), and thus will not reflect the accuracy of translated patches.

Moreover, FID is a highly biased estimator \cite{Binkowski2018}, and has been shown to not be correlated with classification accuracy \cite{Ravuri2019SeeingIN}, as is the concern of this work.

\subsubsection{Noise}

Since both directions of translation (PAS to target and vice versa) obtain visually good results, it is unclear why the performance of MDS1 and MDS2 vary so much. Recent literature \cite{Bashkirova19} indicates that cycle-consistency $\mathcal{L}_\text{cyc}$, Eq.\ \eqref{eq:CycleGAN_loss}, forces image-to-image translation models to hide information necessary for proper reconstruction of $A_\text{rec}$ and $B_\text{rec}$ in the form of imperceptible low amplitude, high frequency noise in $A'$ and $B'$ (see Figure \ref{fig:stain_transfer_diagrams}). It is reasonable to assume that such noise will have an affect on the model's performance.

\begin{figure*}[ht]
	\begin{center}
		\settoheight{\tempdima}{\includegraphics[width=0.105\textwidth]{img/IFTA_Nx_0010_03_glomeruli_patch_59_orig.png}}%
		\begin{tabular}{ l@{ }c@{ }c@{ }c@{ }c@{ }c@{ }c@{ }c@{ }c@{ }}
			{ }& Jones &CD68 & Sirius Red & CD34 & Jones &CD68 & Sirius Red & CD34 
			\\
			\rowname{0} &
			\includegraphics[width=0.10\textwidth]{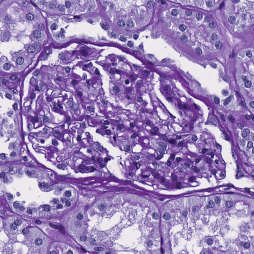} &
			\includegraphics[width=0.10\textwidth]{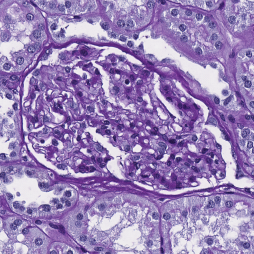} &
			\includegraphics[width=0.10\textwidth]{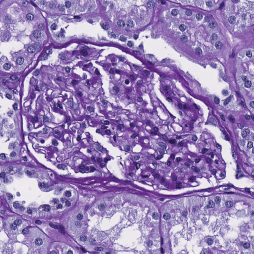} &
			\includegraphics[width=0.10\textwidth]{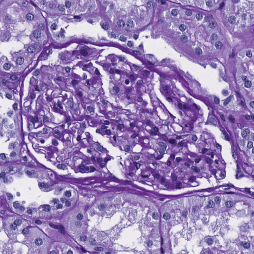} & { }
			\includegraphics[width=0.10\textwidth]{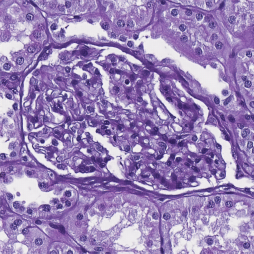} &
			\includegraphics[width=0.10\textwidth]{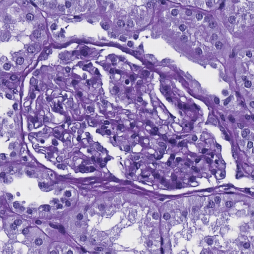} &
			\includegraphics[width=0.10\textwidth]{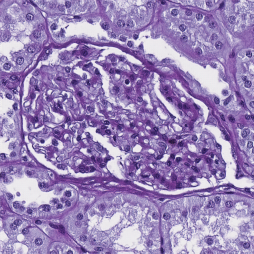} &
			\includegraphics[width=0.10\textwidth]{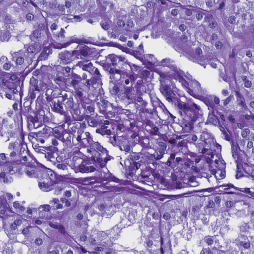}

			\\[-0.5ex]
			\rowname{0.1} &
			\includegraphics[width=0.10\textwidth]{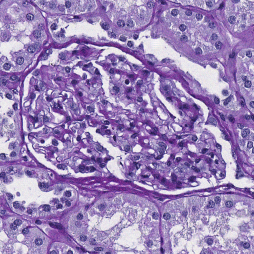} &
			\includegraphics[width=0.10\textwidth]{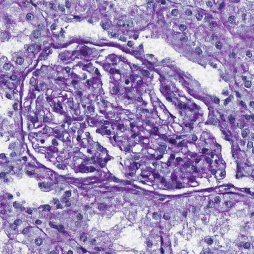} &
			\includegraphics[width=0.10\textwidth]{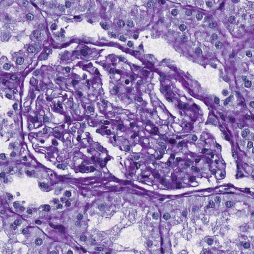} &
			\includegraphics[width=0.10\textwidth]{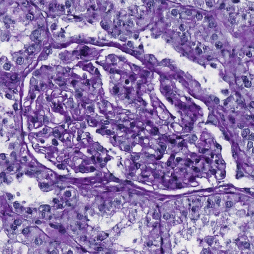} & { }
			\includegraphics[width=0.10\textwidth]{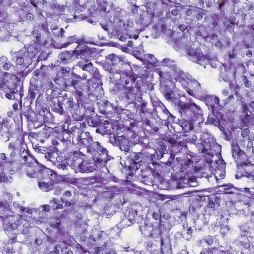} &
			\includegraphics[width=0.10\textwidth]{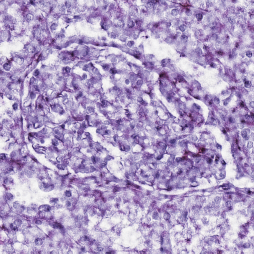} &
			\includegraphics[width=0.10\textwidth]{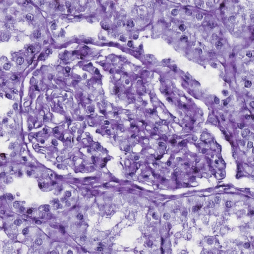} &
			\includegraphics[width=0.10\textwidth]{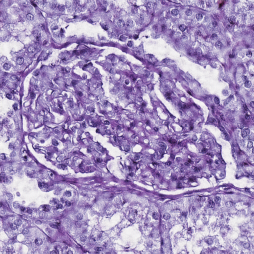}

			\\[-0.5ex]
			
			\rowname{0.5} &
			\includegraphics[width=0.10\textwidth]{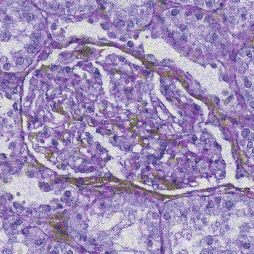} &
			\includegraphics[width=0.10\textwidth]{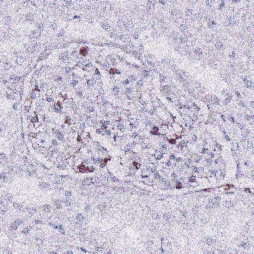} &
			\includegraphics[width=0.10\textwidth]{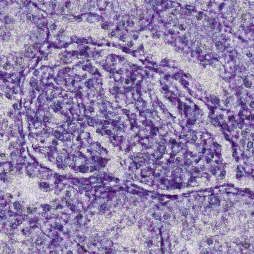} &
			\includegraphics[width=0.10\textwidth]{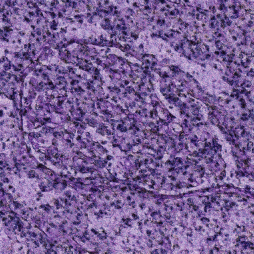} & { }
			\includegraphics[width=0.10\textwidth]{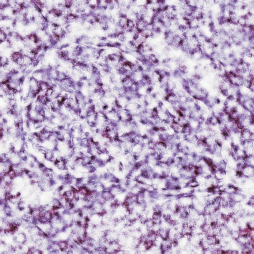} &
			\includegraphics[width=0.10\textwidth]{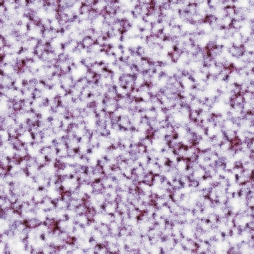} &
			\includegraphics[width=0.10\textwidth]{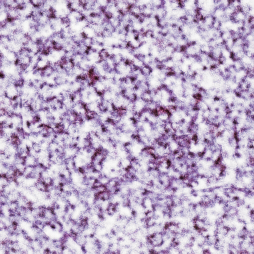} &
			\includegraphics[width=0.10\textwidth]{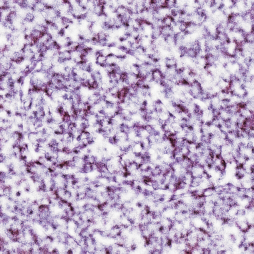} 
			\\[-0.5ex]
			\rowname{0.9} &
			\includegraphics[width=0.10\textwidth]{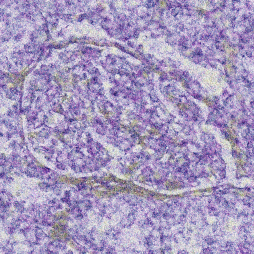} &
			\includegraphics[width=0.10\textwidth]{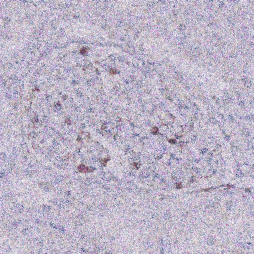} &
			\includegraphics[width=0.10\textwidth]{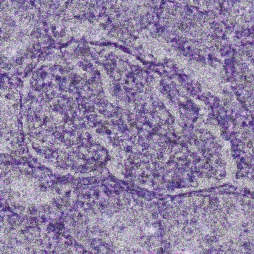} &
			\includegraphics[width=0.10\textwidth]{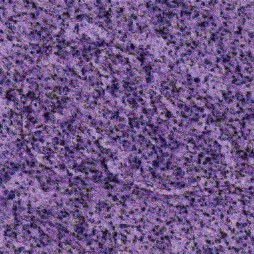} & { }
			\includegraphics[width=0.10\textwidth]{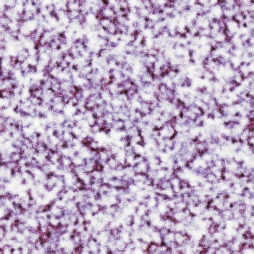} &
			\includegraphics[width=0.10\textwidth]{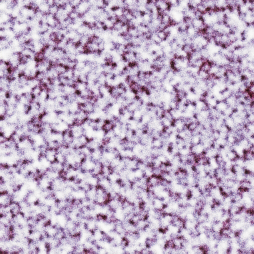} &
			\includegraphics[width=0.10\textwidth]{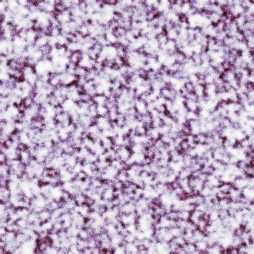} &
			\includegraphics[width=0.10\textwidth]{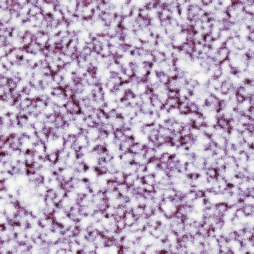}
			\\
			&
			\multicolumn{4}{c}{CycleGAN} & \multicolumn{4}{c}{StarGAN}
		\end{tabular}
		\caption{Effects of additive zero-mean Gaussian noise added to intermediate representation of PAS patch to reconstruction.}
		\label{fig:noisy_reconstructions}
	\end{center}
\end{figure*}

To illustrate this phenomena, we use the composition of translations $\textrm{PAS} \rightarrow \textrm{Target} + \mathcal{N}(0,\sigma) \rightarrow \textrm{PAS}$, where $\mathcal{N}(0,\sigma)$ is a zero mean Gaussian distribution with standard deviation $\sigma$. Figure \ref{fig:noisy_reconstructions} presents reconstructions of the same PAS image after translation to each target stain, with different standard deviation of additive noise. This shows that not all target stain translators encode information in the same way. For example, adding noise with a standard deviation of $0.5$ to the CD68 intermediate stain results in a higher reconstruction error than adding the same noise in the Jones H\&E  intermediate stain.  We suspect that the level of noise in each target staining correlates with the difficulty of translation i.e.\ harder translations require more noise.
StarGAN appears to be more sensitive to the additive noise than CycleGAN. Potentially because it performs the much harder task of multi-stain translation. In order to properly reconstruct its input, more information may need to be hidden and thus it may be more sensitive to additive noise.

In order to quantify this sensitivity to noise, Bashkirova et al.\ \cite{Bashkirova19} propose the following measure: the mean pixel difference $\bar{\delta}$ between the translations with and without noise is measured and is repeated for a range of standard deviations $\sigma$. The area under the resulting $\sigma$ - $\bar{\delta}$ curve summarises the translator's sensitivity (lower is better, i.e.\ less sensitive). This is calculated here using the target-to-source translations of $100$ random PAS patches from each class and $\sigma \in [0,0.025,0.05,0.075,0.1]$, see Table \ref{tab:sensitivity_to_noise}.

\begin{table}[ht]
	\centering
	\small{
		\begin{tabular}{L{1.425cm} L{1.5cm} L{1.2cm} L{1.45cm} L{1.2cm}}
		    \hline \\[-2ex]
	        & {Jones H\&E} & {CD68} & {Sirius Red} & {CD34} \\
			\hhline{=====} \\[-2ex]
			CycleGAN & 5.359 & 14.778 & 7.423 & 11.841 \\ 
			StarGAN & 11.672 & 43.602 & 17.28  & 16.288 \\
			\hline
		\end{tabular}
	}
	\caption{Sensitivity to noise calculated for each target stain based on 100 random patches from each class.}
	\label{tab:sensitivity_to_noise}
\end{table}

These were derived from the same PAS source patches but the generators were trained on different target data. Some caution should therefore be taken when comparing between stains, however, these are as close as possible to being comparable. The orders of sensitivity correlate with both the MDS1 and MDS\textasteriskcentered1 $F_1$ scores presented in Table \ref{tab: quantitative results}.
It is impossible to do the same analysis for MDS2, i.e.\ $\textrm{Target}  \rightarrow \textrm{PAS} \rightarrow \textrm{Target}$, as the target datasets are different.

The relatively high sensitivity to noise  (therefore potentially high levels of noise present in the translation) offers an explanation for UDA-CGAN's relatively low precision in CD68 and CD34, and UDA-\textasteriskcentered GAN's relatively low precision overall.

\begin{figure}[!ht]
	\begin{center}
	\small{
		\settoheight{\tempdima}{\includegraphics[width=0.105\textwidth]{img/IFTA_Nx_0010_03_glomeruli_patch_59_orig.png}}%
		\begin{tabular}{ c@{ }c@{ }c@{ }c@{ }}
			Sirius Red & PAS Seg.\ & CD34  & PAS Seg.\
			\\
			\rowname{Original} 
			\includegraphics[width=0.10\textwidth]{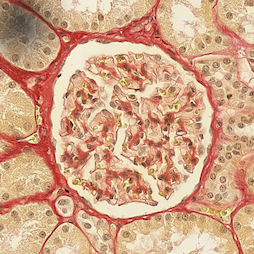} & & \includegraphics[width=0.10\textwidth]{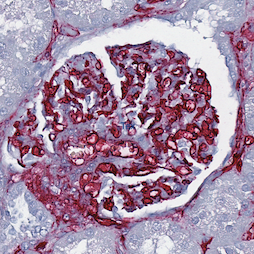}
			\\
			\rowname{1$^{\text{st}}$ Epoch} 
			\includegraphics[width=0.10\textwidth]{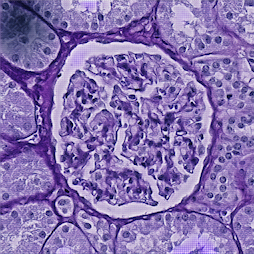} & \includegraphics[width=0.10\textwidth]{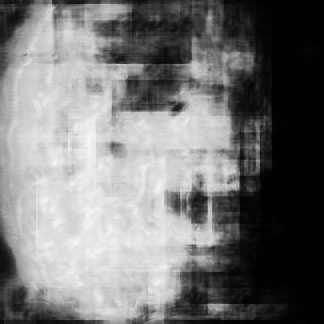} &  \includegraphics[width=0.10\textwidth]{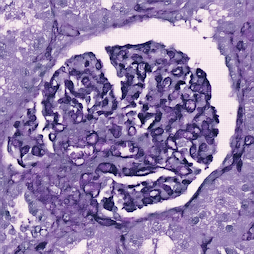} & \includegraphics[width=0.10\textwidth]{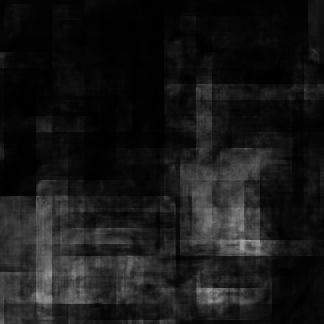} 
			\\
			& \footnotesize{$0.66$} & &  \footnotesize{$0.00$} 
			\\
			\rowname{10$^{\text{th}}$ Epoch} 
			\includegraphics[width=0.10\textwidth]{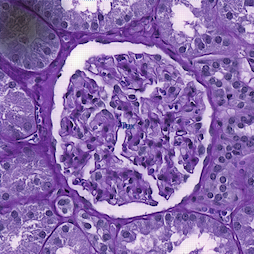} &
			\includegraphics[width=0.10\textwidth]{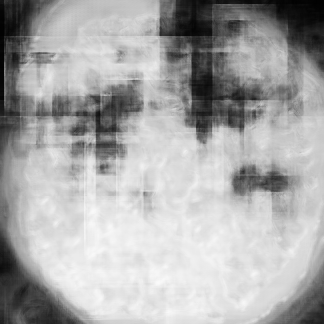} &
			\includegraphics[width=0.10\textwidth]{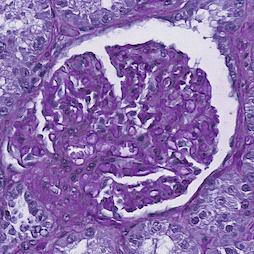} &
			\includegraphics[width=0.10\textwidth]{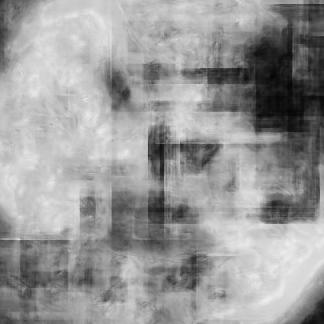}
			\\
			& \footnotesize{$0.71$} & & \footnotesize{$0.65$}
			\\
			\rowname{20$^{\text{th}}$ Epoch} 
			\includegraphics[width=0.10\textwidth]{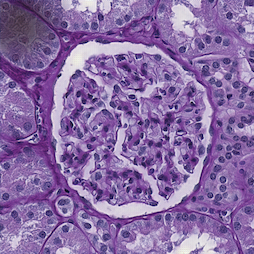} &
			\includegraphics[width=0.10\textwidth]{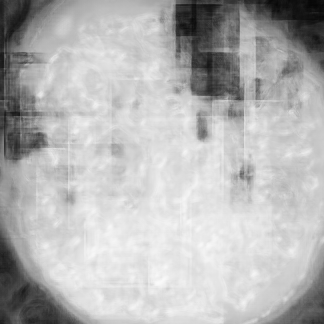} &
			\includegraphics[width=0.10\textwidth]{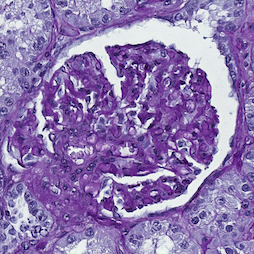} &
			\includegraphics[width=0.10\textwidth]{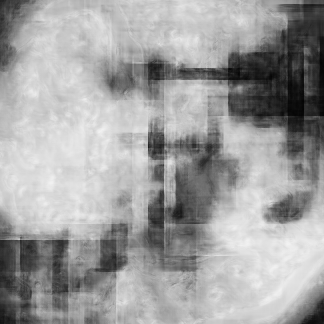}
			\\
			& \footnotesize{$0.76$} & &
			\footnotesize{$0.65$}
			\\
			\rowname{30$^{\text{th}}$ Epoch} 
			\includegraphics[width=0.10\textwidth]{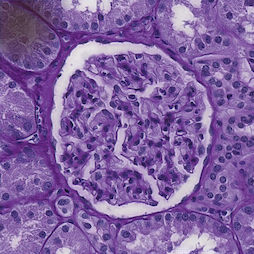} &
			\includegraphics[width=0.10\textwidth]{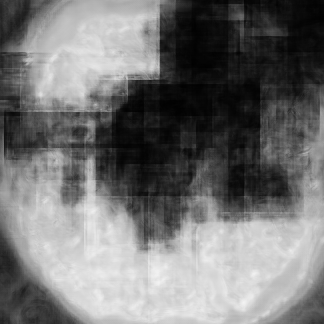} &
			\includegraphics[width=0.10\textwidth]{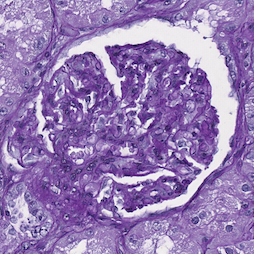} &
			\includegraphics[width=0.10\textwidth]{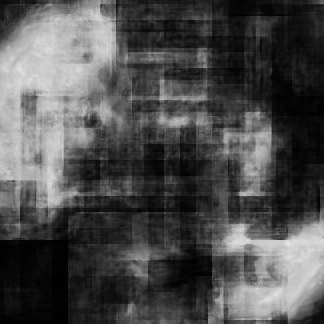}
			\\
			& \footnotesize{$0.43$} & & \footnotesize{$0.28$}
			\\
			\rowname{40$^{\text{th}} Epoch$} 
			\includegraphics[width=0.10\textwidth]{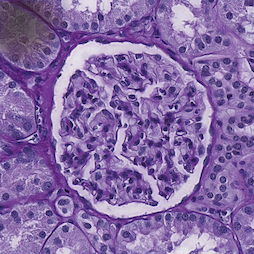} &
			\includegraphics[width=0.10\textwidth]{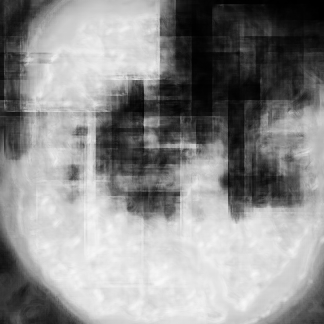} &
			\includegraphics[width=0.10\textwidth]{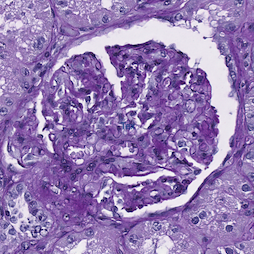} &
			\includegraphics[width=0.10\textwidth]{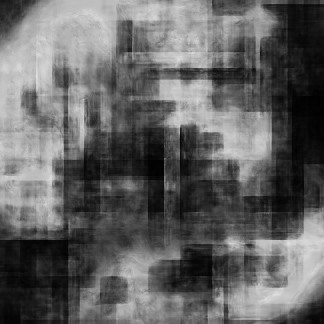}
			\\
			& \footnotesize{$0.59$} &  & \footnotesize{$0.33$}
			\\
			\rowname{50$^{\text{th}}$ Epoch} 
			\includegraphics[width=0.10\textwidth]{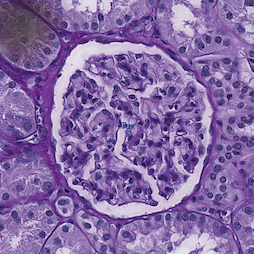} &
			\includegraphics[width=0.10\textwidth]{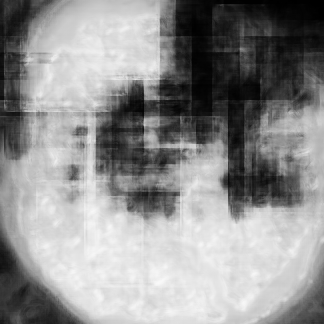} &
			\includegraphics[width=0.10\textwidth]{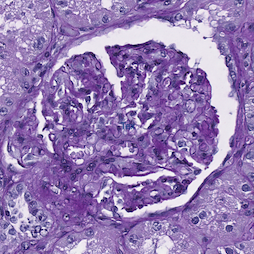} &
			\includegraphics[width=0.10\textwidth]{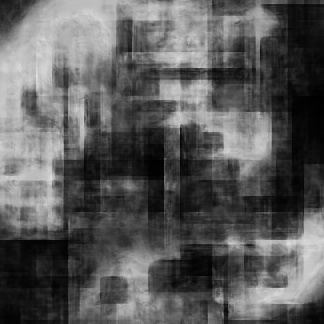}
			\\
			& \footnotesize{$0.61$} & & \footnotesize{$0.29$}\\
		\end{tabular}
		}
		\caption{Glomeruli patches translated to PAS using CycleGAN models from different training epochs with PAS glomeruli segmentations (i.e.\ MDS1 approach) and their F$_1$ scores. N.B.\ images look realistic but segmentations vary widely.}
		\label{fig:cyclegan_epochs_seg}
	\end{center}
\end{figure}

For further illustration, Figure \ref{fig:cyclegan_epochs_seg} presents a Sirius Red and CD34 glomerulus patch translated to PAS using translators taken from different CycleGAN training epochs, along with its segmentation using the overall best performing PAS model (which, being trained on PAS, is sensitive to its distribution). Unintuitively, good visual appearance does not correlate with reliable segmentation (this holds for all stains), which  suggests that there is a limitation on the direct application of cycle-constrained image-to-image translation methods in clinical practice. Brieu et al. \cite{Brieu2019DomainAA} chose the set of training epochs by visual inspection, however, these findings indicate that this is not a good strategy and may lead to the model learning irrelevant features. Furthermore, weakly-supervised approaches based on the performance of an already trained model may not be a good alternative, as the learned features may differ between models and are not necessary task-related.

These analyses, and recent studies on the adversarial nature of CycleGANs \cite{Chu2017CycleGANAM,Wolterink2020DeepLG}, lead to the hypothesis that the translations suffer from invisible artifacts produced by the CycleGAN. The extent and type of these artifacts could be related to stain differences. Stains with a greater difference require more complicated translation, forcing the translators to hallucinate specific features, as confirmed by Mercan et al.\ \cite{Mercan2020VirtualSF}. From this perspective, it can be understood why UDA-CGAN outperforms MDS1 and MDS2 as the model is forced to be robust to artifacts produced by different translations. In this sense, the lack of improvement offered by Multi UDA-CGAN model (see Table \ref{tab: quantitative results}) over UDA-CGAN can be understood. In MDS1, these artifacts hamper the performances as translated images could act as adversarial examples, while in MDS2, these artifacts could be considered by the model as features. \color{black} We hypothesise that a mechanism for assessing the quality of the translation, such as FID \cite{FID_Score} in natural images, would offer further improvement.\color{black}

This only goes to emphasise the care and consideration needed for the direct application of unpaired image-to-image translations in the medical domain, and when used in clinical practice the implications should be held up to even greater scrutiny. Since the proposed approach merely uses the translations to remove a model's bias, see Figure \ref{fig:tsne_rgb_vs_gan}, we largely avoid such concerns.
\color{black}

\section{Conclusions}

\label{sec:conclusions}

To summarise, this article presents a state-of-the-art model, UDA-GAN, that outperforms all other existing pixel-space alignment approaches in five different stainings. The model is domain invariant \color{black} (including to unseen stains)\color{black}, can be easily extended to new stainings, and the training procedure is general, in that it can be used in different segmentation and classification tasks.

The usefulness of ``translating'' stainings has therefore been demonstrated in this context but it should be emphasised that diagnostically relevant information can be lost on the way. We should also emphasise that the primary intention of this translation is to save on resources and time for annotation, not to replace the process of physical staining modalities.

UDA-GAN's stain invariance has been shown through quantitative results and by analysing its feature distribution and attention, which demonstrate that patches from all stainings are more aligned than with competing approaches. The approach uses pixel space alignment, which aids in visual interpretation and verification. 
The results have been discussed and related to those found in the literature. Namely that choosing translation model epochs by visual inspection is not the best approach (although no formal method for selecting the epochs to be used nor evaluating quality exists) and that the direction of translation of the data cannot be simply prescribed a priori. Finally, UDA-GANs limitations were presented; it is still unclear exactly why the model's performance degrades with certain stainings despite visually accurate translations. We hypothesise that some stainings are far from each other in terms of biological structures highlighted, in particular immunohistochemical staining methods highlighting migratory immune cells can be far from conventional histology staining methods making the translation task harder. \color{black} Therefore, the translation model is forced to introduce hidden information. 

Although UDA-GAN approaches diminish these effects to some extent, all these findings indicate that the application of image-to-image translation to medical imaging should be done with careful consideration. \color{black}

\section*{Acknowledgements}
This work was supported by: ERACoSysMed and e:Med initiatives by the German Ministry of Research and Education (BMBF); SysMIFTA (project management PTJ, FKZ 031L-0085A; Agence National de la Recherche, ANR, project number ANR-15—CMED-0004); SYSIMIT (project management DLR, FKZ 01ZX1608A); and the French Government through co-tutelle PhD funding.
We thank Nvidia Corporation for donating a Quadro P6000 GPU and the \emph{Centre de Calcul de l'Université de Strasbourg} for access to the GPUs used for this research.
We also thank the MHH team for providing high-quality images and annotations, specifically Nicole Kroenke for excellent technical assistance, Nadine Schaadt for image management and quality control, and Valery Volk and Jessica Schmitz for annotations under the supervision of domain experts.

\bibliographystyle{unsrt} 
\bibliography{main} 
\end{document}